%% file: taas25.tex
\documentclass[manuscript,screen]{acmart}
\AtBeginDocument{%
  }

\setcopyright{acmlicensed}
\copyrightyear{2025}
\acmYear{2025}
\acmDOI{XXXXXXX.XXXXXXX}
\acmConference[Conference acronym 'XX]{Make sure to enter the correct
  conference title from your rights confirmation email}{June 03--05,
  2018}{Woodstock, NY}
\acmISBN{978-1-4503-XXXX-X/2018/06}

\usepackage{comment}
\usepackage{csquotes}
\usepackage{tikz}
\tikzset{every picture/.style={font=\small}}
\usetikzlibrary{calc,positioning,backgrounds,arrows,shapes,fit,shadows,arrows.meta}
\pgfdeclarelayer{midground}
\pgfsetlayers{background,midground,main}
\usepackage{enumitem}
\usepackage{xspace}
\usepackage[nolist]{acronym}
\usepackage[theorems, skins, breakable, listings]{tcolorbox}
\usepackage{footnote}
\usepackage{todonotes}
\usepackage{subcaption}

\usepackage{tabularx}
\newcolumntype{Y}{>{\raggedleft\arraybackslash}X}
\newcolumntype{Z}{>{\centering\let\newline\\\arraybackslash\hspace{0pt}}X}
\newcolumntype{H}{>{\setbox0=\hbox\bgroup}c<{\egroup}@{}}
\usepackage{colortbl}
\newsavebox\CBox
\def\textBF#1{\sbox\CBox{#1}\resizebox{\wd\CBox}{\ht\CBox}{\textbf{#1}}}
\usepackage{multirow}
\input{KIT_colors}

\newtcbtheorem[use counter=prompt]{prompt}{Prompt}{%
	attach title to upper,%
	after title={\ },%
	colback=white,%
	colframe=kit-blue100,%
	fonttitle={\bfseries\color{black}},%
	if odd page={
			sharp corners=west,%
			rightrule=3mm,%
		}{
			sharp corners=east,%
			leftrule=3mm,%
		},%
	subtitle style={%
			boxrule=0.4pt,%
			colback=kit-blue70,%
			fonttitle={\bfseries\color{white}}%
		}%
}{prompt}

\BeforeBeginEnvironment{prompt}{\savenotes}
\AfterEndEnvironment{prompt}{\spewnotes}

\newtcolorbox{conclusion}{%
	attach title to upper={:\ },%
	colback=white,%
	colframe=kit-green100,%
	fonttitle={\bfseries\color{black}},%
	if odd page={
			sharp corners=west,%
			rightrule=3mm,%
		}{
			sharp corners=east,%
			leftrule=3mm,%
		},%
	subtitle style={%
			boxrule=0.4pt,%
			colback=kit-green70,%
			fonttitle={\bfseries\color{white}}%
		}%
}

\DeclareMathOperator{\FP}{FP}
\DeclareMathOperator{\TP}{TP}

\DeclareMathOperator{\FN}{FN}

\usepackage{pifont}
\newcommand{\cmark}{\ding{51}}%
\newcommand{\xmark}{\ding{55}}%

\begin{document}

\input{abbrev.tex}
\renewcommand{\sectionautorefname}{Section}
\renewcommand{\subsectionautorefname}{Section}
\renewcommand{\subsubsectionautorefname}{Section}

\title{Who's Who? LLM-assisted Software Traceability with Architecture Entity Recognition}


\author{Dominik Fuchß}
\orcid{0000-0001-6410-6769}
\email{dominik.fuchss@kit.edu}
\affiliation{
  \institution{Karlsruhe Institute of Technology (KIT)}
  \city{Karlsruhe}
  \country{Germany}
}

\author{Haoyu Liu}
\orcid{0009-0002-7676-5010}
\email{haoyu.liu@kit.edu}
\affiliation{
  \institution{Karlsruhe Institute of Technology (KIT)}
  \city{Karlsruhe}
  \country{Germany}
}

\author{Sophie Corallo}
\orcid{0000-0002-1531-2977}
\email{sophie.corallo@kit.edu}
\affiliation{
  \institution{Karlsruhe Institute of Technology (KIT)}
  \city{Karlsruhe}
  \country{Germany}
}

\author{Tobias Hey}
\orcid{0000-0003-0381-1020}
\email{hey@kit.edu}
\affiliation{
  \institution{Karlsruhe Institute of Technology (KIT)}
  \city{Karlsruhe}
  \country{Germany}
}

\author{Jan Keim}
\orcid{0000-0002-8899-7081}
\email{jan.keim@kit.edu}
\affiliation{
  \institution{Karlsruhe Institute of Technology (KIT)}
  \city{Karlsruhe}
  \country{Germany}
}

\author{Johannes von Geisau}
\orcid{0009-0003-9736-8150}
\email{uiwyw@student.kit.edu}
\affiliation{
  \institution{Karlsruhe Institute of Technology (KIT)}
  \city{Karlsruhe}
  \country{Germany}
}

\author{Anne Koziolek}
\orcid{0000-0002-1593-3394}
\email{koziolek@kit.edu}
\affiliation{
  \institution{Karlsruhe Institute of Technology (KIT)}
  \city{Karlsruhe}
  \country{Germany}
}


\begin{abstract}
Identifying architecturally relevant entities in textual artifacts is crucial for Traceability Link Recovery (TLR) between Software Architecture Documentation (SAD) and source code.
While Software Architecture Models (SAMs) can bridge the semantic gap between these artifacts, their manual creation is time-consuming.
Large Language Models (LLMs) offer new capabilities for extracting architectural entities from SAD and source code to construct SAMs automatically or establish direct trace links.
This paper extends our ICSA~2025 paper~\cite{fuchss_enabling_2025}, which introduced \ExArch for LLM-based architecture component name extraction.
The extension contributes the novel \ArTEMiS approach, an extended evaluation with additional \LLMs, configurations, a revised benchmark, and a combined evaluation of both approaches.
Specifically, this paper presents the following approaches:
ExArch extracts component names as simple SAMs from SAD and source code to eliminate the need for manual SAM creation, while ArTEMiS identifies architectural entities in documentation and matches them with (manually or automatically generated) SAM entities.
Our evaluation compares against state-of-the-art approaches SWATTR, TransArC and ArDoCode.
TransArC achieves strong performance (F1: 0.87) but requires manually created SAMs; ExArch achieves comparable results (F1: 0.86) using only SAD and code.
ArTEMiS is on par with the traditional heuristic-based SWATTR (F1: 0.81) and can successfully replace it when integrated with TransArC.
The combination of ArTEMiS and ExArch outperforms ArDoCode, the best baseline without manual SAMs.
Our results demonstrate that LLMs can effectively identify architectural entities in textual artifacts, enabling automated SAM generation and TLR, making architecture-code traceability more practical and accessible.
\end{abstract}




\maketitle

\input{sections/intro.tex}
\input{sections/rw.tex}
\input{sections/approach.tex}
\input{sections/eval.tex}
\input{sections/conclusion.tex}

\section*{Data Availability Statement}
We provide our code, baselines, evaluation data, and results in our replication packages~\cite{replication,replication_taas25}.

\begin{acks}
This work was funded by Core Informatics at KIT (KiKIT) of the Helmholtz Assoc. (HGF), the German Research Foundation (DFG) - SFB 1608 - 501798263, the Topic Engineering Secure Systems of the Helmholtz Association (HGF), and supported by KASTEL Security Research Labs, Karlsruhe.
\end{acks}

\bibliographystyle{ACM-Reference-Format}
\bibliography{taas25}

\appendix
\input{sections/appendix}

\end{document}

%% file: KIT_colors.tex
\definecolor{kit-green}{RGB}{0, 150, 130}
\definecolor{kit-green100}{RGB}{0, 150, 130}
\definecolor{kit-green90}{rgb}{0.1, 0.6294, 0.5588}
\definecolor{kit-green80}{rgb}{0.2, 0.6706, 0.6078}
\definecolor{kit-green75}{rgb}{0.25, 0.6912, 0.6324}
\definecolor{kit-green70}{rgb}{0.3, 0.7118, 0.6569}
\definecolor{kit-green60}{rgb}{0.4, 0.7529, 0.7059}
\definecolor{kit-green50}{rgb}{0.5, 0.7941, 0.7549}
\definecolor{kit-green40}{rgb}{0.6, 0.8353, 0.8039}
\definecolor{kit-green30}{rgb}{0.7, 0.8765, 0.8529}
\definecolor{kit-green25}{rgb}{0.75, 0.8971, 0.8775}
\definecolor{kit-green20}{rgb}{0.8, 0.9176, 0.902}
\definecolor{kit-green15}{rgb}{0.85, 0.9382, 0.9265}
\definecolor{kit-green10}{rgb}{0.9, 0.9588, 0.951}
\definecolor{kit-green5}{rgb}{0.95, 0.9794, 0.9755}

\definecolor{kit-blue}{RGB}{70, 100, 170}
\definecolor{kit-blue100}{RGB}{70, 100, 170}
\definecolor{kit-blue90}{rgb}{0.3471, 0.4529, 0.7}
\definecolor{kit-blue80}{rgb}{0.4196, 0.5137, 0.7333}
\definecolor{kit-blue75}{rgb}{0.4559, 0.5441, 0.75}
\definecolor{kit-blue70}{rgb}{0.4922, 0.5745, 0.7667}
\definecolor{kit-blue60}{rgb}{0.5647, 0.6353, 0.8}
\definecolor{kit-blue50}{rgb}{0.6373, 0.6961, 0.8333}
\definecolor{kit-blue40}{rgb}{0.7098, 0.7569, 0.8667}
\definecolor{kit-blue30}{rgb}{0.7824, 0.8176, 0.9}
\definecolor{kit-blue25}{rgb}{0.8186, 0.848, 0.9167}
\definecolor{kit-blue20}{rgb}{0.8549, 0.8784, 0.9333}
\definecolor{kit-blue15}{rgb}{0.8912, 0.9088, 0.95}
\definecolor{kit-blue10}{rgb}{0.9275, 0.9392, 0.9667}
\definecolor{kit-blue5}{rgb}{0.9637, 0.9696, 0.9833}

\definecolor{kit-red}{RGB}{162, 34, 35}
\definecolor{kit-red100}{RGB}{162, 34, 35}
\definecolor{kit-red90}{rgb}{0.6718, 0.22, 0.2235}
\definecolor{kit-red80}{rgb}{0.7082, 0.3067, 0.3098}
\definecolor{kit-red75}{rgb}{0.7265, 0.35, 0.3529}
\definecolor{kit-red70}{rgb}{0.7447, 0.3933, 0.3961}
\definecolor{kit-red60}{rgb}{0.7812, 0.48, 0.4824}
\definecolor{kit-red50}{rgb}{0.8176, 0.5667, 0.5686}
\definecolor{kit-red40}{rgb}{0.8541, 0.6533, 0.6549}
\definecolor{kit-red30}{rgb}{0.8906, 0.74, 0.7412}
\definecolor{kit-red25}{rgb}{0.9088, 0.7833, 0.7843}
\definecolor{kit-red20}{rgb}{0.9271, 0.8267, 0.8275}
\definecolor{kit-red15}{rgb}{0.9453, 0.87, 0.8706}
\definecolor{kit-red10}{rgb}{0.9635, 0.9133, 0.9137}
\definecolor{kit-red5}{rgb}{0.9818, 0.9567, 0.9569}

\definecolor{kit-yellow}{RGB}{252, 229, 0}
\definecolor{kit-yellow100}{RGB}{252, 229, 0}
\definecolor{kit-yellow90}{rgb}{0.9894, 0.9082, 0.1}
\definecolor{kit-yellow80}{rgb}{0.9906, 0.9184, 0.2}
\definecolor{kit-yellow75}{rgb}{0.9912, 0.9235, 0.25}
\definecolor{kit-yellow70}{rgb}{0.9918, 0.9286, 0.3}
\definecolor{kit-yellow60}{rgb}{0.9929, 0.9388, 0.4}
\definecolor{kit-yellow50}{rgb}{0.9941, 0.949, 0.5}
\definecolor{kit-yellow40}{rgb}{0.9953, 0.9592, 0.6}
\definecolor{kit-yellow30}{rgb}{0.9965, 0.9694, 0.7}
\definecolor{kit-yellow25}{rgb}{0.9971, 0.9745, 0.75}
\definecolor{kit-yellow20}{rgb}{0.9976, 0.9796, 0.8}
\definecolor{kit-yellow15}{rgb}{0.9982, 0.9847, 0.85}
\definecolor{kit-yellow10}{rgb}{0.9988, 0.9898, 0.9}
\definecolor{kit-yellow5}{rgb}{0.9994, 0.9949, 0.95}

\definecolor{kit-orange}{RGB}{223, 155, 27}
\definecolor{kit-orange100}{RGB}{223, 155, 27}
\definecolor{kit-orange90}{rgb}{0.8871, 0.6471, 0.1953}
\definecolor{kit-orange80}{rgb}{0.8996, 0.6863, 0.2847}
\definecolor{kit-orange75}{rgb}{0.9059, 0.7059, 0.3294}
\definecolor{kit-orange70}{rgb}{0.9122, 0.7255, 0.3741}
\definecolor{kit-orange60}{rgb}{0.9247, 0.7647, 0.4635}
\definecolor{kit-orange50}{rgb}{0.9373, 0.8039, 0.5529}
\definecolor{kit-orange40}{rgb}{0.9498, 0.8431, 0.6424}
\definecolor{kit-orange30}{rgb}{0.9624, 0.8824, 0.7318}
\definecolor{kit-orange25}{rgb}{0.9686, 0.902, 0.7765}
\definecolor{kit-orange20}{rgb}{0.9749, 0.9216, 0.8212}
\definecolor{kit-orange15}{rgb}{0.9812, 0.9412, 0.8659}
\definecolor{kit-orange10}{rgb}{0.9875, 0.9608, 0.9106}
\definecolor{kit-orange5}{rgb}{0.9937, 0.9804, 0.9553}

\definecolor{kit-lightgreen}{RGB}{140, 182, 60}
\definecolor{kit-lightgreen100}{RGB}{140, 182, 60}
\definecolor{kit-lightgreen90}{rgb}{0.5941, 0.7424, 0.3118}
\definecolor{kit-lightgreen80}{rgb}{0.6392, 0.771, 0.3882}
\definecolor{kit-lightgreen75}{rgb}{0.6618, 0.7853, 0.4265}
\definecolor{kit-lightgreen70}{rgb}{0.6843, 0.7996, 0.4647}
\definecolor{kit-lightgreen60}{rgb}{0.7294, 0.8282, 0.5412}
\definecolor{kit-lightgreen50}{rgb}{0.7745, 0.8569, 0.6176}
\definecolor{kit-lightgreen40}{rgb}{0.8196, 0.8855, 0.6941}
\definecolor{kit-lightgreen30}{rgb}{0.8647, 0.9141, 0.7706}
\definecolor{kit-lightgreen25}{rgb}{0.8873, 0.9284, 0.8088}
\definecolor{kit-lightgreen20}{rgb}{0.9098, 0.9427, 0.8471}
\definecolor{kit-lightgreen15}{rgb}{0.9324, 0.9571, 0.8853}
\definecolor{kit-lightgreen10}{rgb}{0.9549, 0.9714, 0.9235}
\definecolor{kit-lightgreen5}{rgb}{0.9775, 0.9857, 0.9618}

\definecolor{kit-purple}{RGB}{163, 16, 124}
\definecolor{kit-purple100}{RGB}{163, 16, 124}
\definecolor{kit-purple90}{rgb}{0.6753, 0.1565, 0.5376}
\definecolor{kit-purple80}{rgb}{0.7114, 0.2502, 0.589}
\definecolor{kit-purple75}{rgb}{0.7294, 0.2971, 0.6147}
\definecolor{kit-purple70}{rgb}{0.7475, 0.3439, 0.6404}
\definecolor{kit-purple60}{rgb}{0.7835, 0.4376, 0.6918}
\definecolor{kit-purple50}{rgb}{0.8196, 0.5314, 0.7431}
\definecolor{kit-purple40}{rgb}{0.8557, 0.6251, 0.7945}
\definecolor{kit-purple30}{rgb}{0.8918, 0.7188, 0.8459}
\definecolor{kit-purple25}{rgb}{0.9098, 0.7657, 0.8716}
\definecolor{kit-purple20}{rgb}{0.9278, 0.8125, 0.8973}
\definecolor{kit-purple15}{rgb}{0.9459, 0.8594, 0.9229}
\definecolor{kit-purple10}{rgb}{0.9639, 0.9063, 0.9486}
\definecolor{kit-purple5}{rgb}{0.982, 0.9531, 0.9743}

\definecolor{kit-brown}{RGB}{167, 130, 46}
\definecolor{kit-brown100}{RGB}{167, 130, 46}
\definecolor{kit-brown90}{rgb}{0.6894, 0.5588, 0.2624}
\definecolor{kit-brown80}{rgb}{0.7239, 0.6078, 0.3443}
\definecolor{kit-brown75}{rgb}{0.7412, 0.6324, 0.3853}
\definecolor{kit-brown70}{rgb}{0.7584, 0.6569, 0.4263}
\definecolor{kit-brown60}{rgb}{0.7929, 0.7059, 0.5082}
\definecolor{kit-brown50}{rgb}{0.8275, 0.7549, 0.5902}
\definecolor{kit-brown40}{rgb}{0.862, 0.8039, 0.6722}
\definecolor{kit-brown30}{rgb}{0.8965, 0.8529, 0.7541}
\definecolor{kit-brown25}{rgb}{0.9137, 0.8775, 0.7951}
\definecolor{kit-brown20}{rgb}{0.931, 0.902, 0.8361}
\definecolor{kit-brown15}{rgb}{0.9482, 0.9265, 0.8771}
\definecolor{kit-brown10}{rgb}{0.9655, 0.951, 0.918}
\definecolor{kit-brown5}{rgb}{0.9827, 0.9755, 0.959}

\definecolor{kit-cyan}{RGB}{35, 161, 224}
\definecolor{kit-cyan100}{RGB}{35, 161, 224}
\definecolor{kit-cyan90}{rgb}{0.2235, 0.6682, 0.8906}
\definecolor{kit-cyan80}{rgb}{0.3098, 0.7051, 0.9027}
\definecolor{kit-cyan75}{rgb}{0.3529, 0.7235, 0.9088}
\definecolor{kit-cyan70}{rgb}{0.3961, 0.742, 0.9149}
\definecolor{kit-cyan60}{rgb}{0.4824, 0.7788, 0.9271}
\definecolor{kit-cyan50}{rgb}{0.5686, 0.8157, 0.9392}
\definecolor{kit-cyan40}{rgb}{0.6549, 0.8525, 0.9514}
\definecolor{kit-cyan30}{rgb}{0.7412, 0.8894, 0.9635}
\definecolor{kit-cyan25}{rgb}{0.7843, 0.9078, 0.9696}
\definecolor{kit-cyan20}{rgb}{0.8275, 0.9263, 0.9757}
\definecolor{kit-cyan15}{rgb}{0.8706, 0.9447, 0.9818}
\definecolor{kit-cyan10}{rgb}{0.9137, 0.9631, 0.9878}
\definecolor{kit-cyan5}{rgb}{0.9569, 0.9816, 0.9939}

\definecolor{kit-gray}{RGB}{0, 0, 0}
\definecolor{kit-gray100}{RGB}{0, 0, 0}
\definecolor{kit-gray90}{rgb}{0.1, 0.1, 0.1}
\definecolor{kit-gray80}{rgb}{0.2, 0.2, 0.2}
\definecolor{kit-gray75}{rgb}{0.25, 0.25, 0.25}
\definecolor{kit-gray70}{rgb}{0.3, 0.3, 0.3}
\definecolor{kit-gray60}{rgb}{0.4, 0.4, 0.4}
\definecolor{kit-gray50}{rgb}{0.5, 0.5, 0.5}
\definecolor{kit-gray40}{rgb}{0.6, 0.6, 0.6}
\definecolor{kit-gray30}{rgb}{0.7, 0.7, 0.7}
\definecolor{kit-gray25}{rgb}{0.75, 0.75, 0.75}
\definecolor{kit-gray20}{rgb}{0.8, 0.8, 0.8}
\definecolor{kit-gray15}{rgb}{0.85, 0.85, 0.85}
\definecolor{kit-gray10}{rgb}{0.9, 0.9, 0.9}
\definecolor{kit-gray5}{rgb}{0.95, 0.95, 0.95}

%% file: abbrev.tex
\begin{acronym}
    \acro{SWATTR}{SoftWare Architecture Text Trace link Recovery}
    \acro{ArCoTL}{ARchitecture-to-COde Trace Linking}
    \acro{TransArC}{Transitive links for Architecture and Code}
    \acro{ArTEMiS}{Architecture Traceability with Entity Matching via Semantic inference}
    \acro{FTLR}{Fine-grained Traceability Link Recovery}
    \acro{TLR}{traceability link recovery}
    \acro{JSD}{probabilistic Jensen Shannon model divergence}
    \acro{VSM}{Vector Space Model}
    \acro{LSI}{Latent Semantic Indexing}
    \acro{WMD}{Word Mover's Distance}
    \acro{SAD}{Software Architecture Documentation}
    \acro{SAM}{Software Architecture Model}
    \acro{LLM}{Large Language Model}
    \acro{NLP}{Natural Language Processing}
    \acro{NER}{Named Entity Recognition}
    \acro{ExArch}{Extracting Architecture}
\end{acronym}

\newcommand{\SWATTR}{\ac{SWATTR}\xspace}
\newcommand{\ArDoCo}{ArDoCo\xspace}
\newcommand{\ArCoTL}{\ac{ArCoTL}\xspace}
\newcommand{\TransArC}{\ac{TransArC}\xspace}
\newcommand{\ExArch}{\ac{ExArch}\xspace}
\newcommand{\ArTEMiS}{\ac{ArTEMiS}\xspace}
\newcommand{\ArDoCode}{ArDoCode\xspace}
\newcommand{\FTLR}{FTLR\xspace}
\newcommand{\TLR}{\ac{TLR}\xspace}
\newcommand{\LSI}{\ac{LSI}\xspace}
\newcommand{\JSD}{\ac{JSD}\xspace}
\newcommand{\VSM}{\ac{VSM}\xspace}
\newcommand{\VSMs}{\acs{VSM}\xspace}
\newcommand{\WMD}{\ac{WMD}\xspace}
\newcommand{\SAD}{\ac{SAD}\xspace}
\newcommand{\SAM}{\ac{SAM}\xspace}
\newcommand{\SADs}{\acp{SAD}\xspace}
\newcommand{\SAMs}{\acp{SAM}\xspace}
\newcommand{\LLM}{\ac{LLM}\xspace}
\newcommand{\LLMs}{\acp{LLM}\xspace}
\newcommand{\NER}{\ac{NER}\xspace}

\newcommand{\SWA}{software architecture\xspace}
\newcommand{\NL}{natural language\xspace}
\newcommand{\NLP}{\ac{NLP}\xspace}

\newcommand{\FOne}{\FOneTxt}
\newcommand{\FOneTxt}{F\textsubscript{1}\xspace}
\newcommand{\FOneSc}{\FOneTxt-score\xspace}
\newcommand{\TLs}{trace links\xspace}
\newcommand{\TL}{trace link\xspace}
\newcommand{\gs}{gold standard\xspace}
\newcommand{\gss}{gold standards\xspace}

\newcommand{\CoT}{Chain-of-thought\xspace}

\newcommand{\GPTFourOMini}{GPT-4o mini\xspace}
\newcommand{\GPTFourO}{GPT-4o\xspace}
\newcommand{\GPTFourTurbo}{GPT-4 Turbo\xspace}
\newcommand{\GPTFour}{GPT-4\xspace}
\newcommand{\GPTFourOne}{GPT-4.1\xspace}
\newcommand{\GPTFive}{GPT-5\xspace}
\newcommand{\GPTThreeFiveTurbo}{GPT-3.5 Turbo\xspace}
\newcommand{\CodellamaThirteenB}{Codellama 13b\xspace}
\newcommand{\LlamaThreeOneEightB}{Llama 3.1 8b\xspace}
\newcommand{\LlamaThreeOneSeventyB}{Llama 3.1 70b\xspace}

%% file: sections/intro.tex
\section{Introduction}
\label{sec:intro}

In software development, numerous artifacts are produced, each representing different levels of abstraction and addressing distinct aspects of the system.
Challenges for architects and developers arise because the relationships between these artifacts are often unclear, which prevents their effective use and causes challenges in maintenance.
To address this, \TLR techniques are used to establish, maintain, and manage explicit \TLs between artifacts.
Improving software quality is closely tied to the creation and management of \TLs~\cite{rempel17,p_tlr_supports11}.

One challenge in linking artifacts is the semantic gap between different types of artifacts.
Bridging this gap is difficult, and automated approaches often misinterpret the underlying semantics.
To address this, some methods suggest using intermediate artifacts to reduce the semantic gap, making it easier to link related artifacts~\cite{aung_2020,nishikawa_clm_2015,moran_improving_2020,keim_transarc_2024}.
For example, the descriptions in design documentation are semantically closer to design artifacts like \SAMs than to code, and \SAMs, in turn, are closer to code.
Consequently, specialized approaches can more easily establish links between \SADs and \SAMs, or between \SAMs and code~\cite{hayes2007requirements,icsa23_ardoco}.
Based on these insights, transitive approaches like \TransArC~\cite{keim_transarc_2024} have been developed.
\TransArC recovers \TLs between \SADs and source code based on manually created \SAMs as intermediate artifacts.
However, these methods are not always applicable, as intermediate artifacts are often unavailable.

The \TLs across the \SAD-\SAM-Code chain are bi-directional and differ in multiplicity.
Bi-directionality means that links can be traversed in both directions: forward (\SAD~$\rightarrow$~\SAM~$\rightarrow$~Code) to identify which code units implement a given architectural description, and backward (Code~$\rightarrow$~\SAM~$\rightarrow$~\SAD) to identify which architectural descriptions and components are responsible for a given code unit.
Accordingly, developers can, for example, trace the root cause of a failure back to the relevant passages in the \SAD.
The two sub-link types also differ in their multiplicity.
\SAD-\SAM links are many-to-many.
This means, on one hand, a single sentence in the \SAD can mention multiple components.
For example the sentence \enquote{The MediaManagement requests the user token from the UserManagement.} links to both components simultaneously.
On the other hand, a single component of the \SAM may be described across multiple sentences throughout the \SAD.
\SAM-Code links are one-to-many in the forward direction: one architectural component (e.g., \texttt{MediaAccess} or \texttt{UserManagement}) can map to multiple implementing source files.
In the backward direction, a source file usually belongs only to one component, making these links many-to-one from code to \SAM.

We propose two complementary \LLM-based approaches that together enable end-to-end \TLR from \SAD to source code without the need for manually created \SAMs.
Both approaches follow a common design philosophy: rather than delegating the entire \TLR task to an \LLM, they decompose the problem and employ \LLMs only for the subtasks where they excel.
Specifically, they focus on identifying and extracting architectural entities.
For the actual trace link recovery, these new approaches are then providing necessary information to the existing heuristic-based approach \TransArC \cite{keim_transarc_2024}.
Consequently, we can formulate our overarching research question:
\begin{tcolorbox}[colback=white, colframe=black, arc=2pt, boxrule=0.7pt]
    \centering\textbf{RQ0} \emph{How well can \LLMs extract architectural elements to enable and assist in \TLR between \SAD and source code?}
\end{tcolorbox}

First, we introduce \emph{\ExArch}, an approach that uses \LLMs to extract component names from \SAD and/or source code.
Since, in practice, \SAMs are regularly unavailable, we aim to recover \TLs between \SAD and code without the need for manually created \SAMs.
To achieve this, we leverage the strength of \LLMs in understanding natural language and code.
We design an approach that uses \LLMs to recover \SAMs in the form of component names from \SAD and/or code.
Component names contribute the required information to apply \TransArC without manually creating \SAMs.
In doing so, we bridge the semantic gap between \SAD and code and empower \TLR between these artifacts.
We emphasize that the component names define a simple \SAM that provides exactly the information for \TLR.
In contrast to the research field of architecture recovery that aims to recover more detailed architecture models, we focus on the information needed to enable state-of-the-art \SAD-to-code \TLR approaches such as \TransArC.
Consequently, we have the following research questions regarding this approach:
\begin{enumerate}[wide=0pt, leftmargin=30pt, labelindent=0pt, labelwidth=20pt, align=left]
    \item[\textbf{RQ1.1}] Is the performance of architecture TLR with LLM-extracted component names as intermediate artifacts~(\ExArch) comparable to using manually created \SAMs?
    \item[\textbf{RQ1.2}] Does \TransArC with \ExArch perform better than state-of-the-art TLR between SAD and code without \SAMs as intermediates?
    \item[\textbf{RQ1.3}] Is the performance of \TransArC with \ExArch with current open-source \LLMs comparable to the performance of closed-source ones?
    \item[\textbf{RQ1.4}] How does the performance of \TransArC with \ExArch differ when using different artifacts to generate the simple \SAMs?
\end{enumerate}

Second, we introduce \emph{\ArTEMiS}, an approach to recover \TLs between \SAD and \SAM using \LLMs.
The approach uses \LLMs for a special form of Named Entity Recognition for software architecture.
It is a modern alternative to traditional heuristic-based \TLR approaches such as the approach by Keim et al. \cite{keim2021_swattr}.
We compare our new \ArTEMiS approach with existing \TLR approaches for \SAD-\SAM \TLR.
Finally, we combine \ArTEMiS with \TransArC and \ExArch to evaluate the influence of \LLM-based \SAD-\SAM \TLR on the performance of \TransArC.
Consequently, we have the following research questions regarding this approach:
\begin{enumerate}[wide=0pt, leftmargin=30pt, labelindent=0pt, labelwidth=20pt, align=left]
    \item[\textbf{RQ2.1}] Is the \LLM-based \TLR from \SAD to \SAM (\ArTEMiS) better than state-of-the-art heuristic-based \TLR?
    \item[\textbf{RQ2.2}] Does \ArTEMiS improve the performance of \TransArC with and without \ExArch?
\end{enumerate}

This paper extends our ICSA~2025 paper~\cite{fuchss_enabling_2025}, which initially presented \ExArch and presented initial evaluation results.
\ExArch extracts architecture component names for transitive \TLR between \SAD and code, removing the need for manually created \SAMs.
\ArTEMiS is a novel contribution of this extension: 
It extracts architecture entities from \SAD similar to Named Entity Recognition, links them to \SAMs, and provides a modern alternative to classical heuristic-based approaches.
We additionally provide an extended evaluation of \ExArch with a broader range of \LLMs, a revised benchmark, and the combined evaluation of \ArTEMiS and \ExArch with \TransArC.
We provide code, baselines, evaluation data, and results in our replication packages~\cite{replication,replication_taas25}.

We structure the remainder of the paper as follows:
Related work is examined in \autoref{sec:rw}.
\ExArch~\cite{fuchss_enabling_2025} is presented in \autoref{sec:approach}, including a description of how we incorporated \TransArC~\cite{keim_transarc_2024} in \autoref{sec:transarc}.
Moreover, in \autoref{sec:artemis}, we describe the \ArTEMiS approach to extract architecture entities from \SAD and how to link them to \SAM.
In \autoref{sec:exp_design}, we describe our experimental design.
Here, we present our initial experiment on the performance of \ExArch.
Moreover, we show the results of our extended evaluation of \ExArch, \ArTEMiS, and their combination.
Lastly, we conclude this paper in \autoref{sec:conclusion}.

%% file: sections/rw.tex
\section{Related Work \& Foundations}
\label{sec:rw}
In this section, we first discuss the ideas and benefits of transitive links, comparing previous works with ours.
Afterward, we focus on the application of \LLMs for \TLR.
Here, we concentrate on the information extraction ability of \LLMs that also motivates our use of \LLMs.
We also discuss the use of \LLMs for \NER that relates to our \ArTEMiS approach.
Finally, we discuss architecture recovery and its differences from our component name recovery.

\subsection{Transitive Trace Links}
Automated \TLR approaches primarily involve comparing different terms across textual artifacts to find terms referring to the same concept.
Consequently, researchers have leveraged semantic similarity techniques developed within the \NLP field to facilitate this process.

Among those techniques, methods like \VSM and \LSI are commonly used~\cite{antoniol_recovering_2002}.
However, software artifacts exist at various levels of abstraction.
This complexity challenges \NLP models, as they often struggle to handle cross-level artifacts effectively, limiting their accuracy in recovering trace links.
Thus, researchers have explored different techniques to address this challenge.
Approaches include incorporating fine-grained information \cite{hey_improving_2021,hey_requirements_2024}, considering dependencies \cite{kuang_dataDep_2015,panichella_when_2013}, and using enriched vocabularies \cite{gao_using_2023}.
More recently, transitive linking through intermediate artifacts has been shown to be promising in bridging those semantic gaps \cite{rodriguez_leveraging_2021,keim_transarc_2024}.

The underlying idea of transitive links is that intermediate artifacts can help to find implicit tracing relationships \cite{moran_comet_2020}.
An early work by \citet{nishikawa_clm_2015} showed the importance of choosing suitable intermediate artifacts for transitive link recovery.
Their approach focuses on establishing transitive links between two artifacts using a third artifact~\cite{nishikawa_clm_2015}.
It uses \VSM to generate initial trace links between various pairs of artifacts.
These links include pairs between use cases, interaction diagrams, code, and test cases.
The experiment was conducted under various settings, including scenarios with no intermediate artifacts and with different artifacts used as intermediates.
Their findings highlight that suitable intermediate artifacts can significantly influence \TLR performance.
For example, interaction diagrams are better intermediates than use cases when recovering trace links between code and test cases.
Instead of only using \VSM to generate initial links for later transitive linking, various \NLP techniques can be combined \cite{moran_comet_2020,rodriguez_leveraging_2021}.
COMET \cite{moran_comet_2020} uses a Bayesian inference framework to treat the recovery process from a probabilistic view.
It uses multiple similarity scores to estimate the model's parameters. 
In contrast to COMET, \citet{rodriguez_leveraging_2021} first combine multiple scores from different sources into a single score, and then evaluate the link based on the final score. 

So far, previous works mainly focused on traceability in requirements \cite{antoniol_recovering_2002,moran_comet_2020,hey_improving_2021,kuang_dataDep_2015,gao_using_2023,rodriguez_leveraging_2021} or test cases \cite{nishikawa_clm_2015,moran_comet_2020,panichella_when_2013}, leaving architectural traceability \cite{icsa23_ardoco,keim_transarc_2024} less explored.
\TransArC~\cite{keim_transarc_2024} has so far achieved the best performance in the architectural traceability benchmark.
It leverages \SAMs as intermediate artifacts to recover trace links between \SAD and code.
Although it achieves significant improvement, its reliance on the existence of well-maintained \SAMs limits its application to a wider range of projects.

Transitive approaches have shown promising results, but cannot work when the intermediate artifacts are missing.
Therefore, it remains a challenge to leverage transitive links when the intermediate artifacts are not completely present.
To enable transitive trace link recovery of \SAD and source code in more settings, we explore using \LLMs to recover the needed information of a \SAM for \TLR.

\subsection{LLMs for Traceability}
With the rich general knowledge obtained from pretraining, \LLMs have been effective at knowledge-intensive software engineering tasks \cite{p_llm4modeling_edu,p_llm4exercise,p_llm4_refactoring}. 
\LLMs have been used for various architecture and modeling tasks~\cite{p_llm4se_survey,p_GenAI_Manifesto}, including supporting design decision making~\cite{p_icsa_ddgen,p_arch_decision_ecsa},
modeling tasks~\cite{p_llm4modeling1,p_llm4modeling2}, and software architecture analysis and generation~\cite{p_llm_col_arch,p_llm4archgen1}.
In the following, we discuss the role of \LLMs in advancing traceability research.
T-BERT is an early adoption of \LLMs for \TLR in an issue-commit setting \cite{lin_tbert_2021}.
The BERT language model was tested using three variants of neural architectures: Twin, Siamese, and Single.
They demonstrated that T-BERT outperforms previous \VSM and recurrent neural network approaches.

With the rise of decoder-only \LLMs like GPT, prompt phrasing greatly influences the language model's output~\cite{rodriguez_prompt_2023}.
They explored the performance of those \LLMs on recovering links with different types of prompts: classification (two artifacts are linked or not), ranking (rank all related artifacts), and \CoT (recovery step-by-step while giving reasons).
Besides, they showed that \LLMs can understand domain-specific terms from the general knowledge obtained by pre-training.
Motivated by this understanding abilities, we explore \LLMs to extract intermediate structures from architecture documents and source code to help \TLR.

\citet{llm_tlr_secReq2GM} explored \LLMs' zero-shot abilities to trace security-related requirements to goal models.
The prompts are tailored to the Goal-oriented Requirements Language's peculiarities.
Although achieving positive results, the approach's high dependence on the task and data limits its further application to our architectural \TLR task.
\citet{llm_codegen_tlr} considered that using requirements-to-code links helps \LLMs to generate code by iteratively reformatting prompts.
They use the gradients between generated code and requirements sentences to identify which part of the requirement is overlooked.
This overlooking link is later used to reformat the prompt.

\citet{fuchss_lissa_2025} created the LiSSA framework for generic \TLR.
They use retrieval-augmented generation and \LLM-based classification to trace artifacts like \SAD, code, and requirements.
LiSSA has also been applied to inter-requirements traceability \cite{hey_requirements_2025}.
However, for \TLR related to architecture models, they did not outperform the state of the art.

Existing \LLM approaches for \TLR show that \LLMs can have the software knowledge needed for our task.
However, our task differs from those already tackled, or the results of the existing approaches did not reach state-of-the-art, so these approaches can't be directly applied.

\subsection{\acf{NER}}
\NER is a subtask of \NLP that focuses on identifying and classifying key entities in text into predefined categories, such as names of persons, organizations, locations, and more.
In software engineering, \NER is often used to extract relevant entities from Q\&A discussions and bug reports, cybersecurity-related texts, and other software artifacts.
Thus, extracting software entities, like frameworks, libraries, and tools from StackOverflow posts is a common task~\cite{tabassum_2020,ye_ner_2016,sun_2020,Veera_2019,Tang2022}.
\citet{zhou_2020} recognizes bug-specific entities, like GUI elements, network entities, drivers, hardware, etc., in bug reports.
In cybersecurity, \NER is used to extract cybersecurity-related terms, such as vulnerabilities or software entities from security websites, news, or cyber threat intelligence texts \cite{chen_2021,tikhomirov_2020,wu_2020,soltani_2025}.
More rarely, \NER is used to extract software entities from user stories~\cite{herwanto_2024}, requirements~\cite{Malik_2022,das_2023}.
In contrast to the mentioned approaches, our approach focuses on extracting component names from architecture documents.
Thus, the entities are not only domain-specific but more closely tied to the project context, which makes the task even more challenging.


\subsection{Architecture Recovery}
Our approach's generation of component names is conceptually related to software architecture recovery research.
Therefore, we give a brief overview of architecture recovery approaches and describe the differences. In the end, we discuss how our approach differs from previous ones.

Structural information, including system dependencies \cite{codeDepOnRecov} and folder structure, is an important source for architecture recovery.
An early approach, ACCD~\cite{tzerpos2000accd}, uses folder and dependency information gathered through static analysis.
Building upon ACCD, dynamic dependency information has been shown to help recovery in some cases~\cite{dynamic_dep_recovery}.
\citet{codeDepOnRecov} examined the impact of code dependency in detail, finding that symbol dependencies yield better recovery accuracy than include dependencies.
In addition to structural information, textual information can also help.
The primary idea is that artifacts belonging to the same component may have similar variable names.
ARC~\cite{arc} recovers architecture using system concerns, which are application-specific features extracted from software corpora.
\citet{lexicalRecov} studied the impact of six different types of identifier names.
They found it would be better to treat identifiers separately by their types and assign suitable weights than to treat them equally in a large vocabulary.
Structural and semantic information can also be combined.
They can be integrated to build weighted call graphs~\cite{Recov_fuse1}, create recovery patterns~\cite{AMARJEET201796},
and be combined using information fusion models~\cite{RecInfoFusion}.
\citet{deduct_recov} showed architecture could also be recovered deductively.
They start from a reference architecture and iteratively use \LLMs to refine it with implementation details.
Although we share similar ideas with architecture recovery, the purpose and final result are different. 
Our approaches are designed to retrieve the required information to reduce the gap for \TLR between architecture documentation and code, i.e., component names.
The recovered model thus does not serve as a complete, functional artifact.
In contrast, architecture recovery focuses on recovering detailed software architecture for more general purposes, like understanding the system's functionality or performance prediction. 

%% file: sections/approach.tex
\section{ExArch: An Approach to Generate Simple SAMs}
\label{sec:approach}
This section describes \acf{ExArch}, our approach to generate a simple \SAM from \SAD and source code to enable transitive \TLR. The approach was introduced by \citet{fuchss_enabling_2025} at the International Conference on Software Architecture (ICSA) 2025.

\input{figures/approach}
\autoref{fig:approach} provides an overview of \ExArch.
\ExArch consists of two main parts:
generating the component names for a \SAM and recovering trace links.
The input for \ExArch is the \SAD and the software project's source code.
Since the whole source code is typically too large to be used as input for an \LLM, we extract features from it.
We then use prompting strategies to extract and generate component names for the intermediate \SAM.
Using these \SAMs, we apply the \TransArC~\cite{keim_transarc_2024} approach to recover trace links between the \SAD and the source code.
In the following, we describe the individual steps in detail.

\subsection{Feature Extraction for Source Code}
\label{sec:feature_extraction}
As \LLMs typically only accept inputs that are smaller than a project's source code, we only provide extracted features from the source code as input.
Since we aim to recover component names from the source code, we do not need to maintain all the information.

In object-oriented programming languages like Java, the package structure can provide valuable information about the architecture of a system \cite{Sinkala22}.
Thus, we extract the package structure from the source code.
\ExArch is not limited to this feature, but the feature extraction in this paper focuses only on providing a list of non-empty source code packages.

\subsection{Prompting Strategies}
\label{sec:prompts}
This section describes the prompting strategies we use to generate the \SAM.
We consider three modes for this paper:
First, we extract the component names only from the \SAD.
Second, we extract the component names only from the source code.
Third, we incorporate both to generate the simple \SAMs.

\paragraph{Extract Component Names from Documentation}
We use \CoT prompting to generate the simple \SAM from the architecture documentation.
We use two prompts:
\begin{prompt}{Documentation to Architecture (1)}{doc2arch1}
    Your task is to identify the high-level components based on the software architecture documentation. In a first step, you shall elaborate on the following documentation: \{Software Architecture Documentation\}
\end{prompt}
\begin{prompt}{Documentation to Architecture (2)}{doc2arch2}
    Now provide a list that only covers the component names. Omit common prefixes and suffixes in the names in camel case.
    Output format:\newline
    - Name1\newline
    - Name2
\end{prompt}
\autoref{prompt:doc2arch1} queries the \LLM to identify high-level components based on the architecture documentation.
We also instruct the \LLM to elaborate on the architecture documentation.
Thus, the \LLM is not restricted to generate output in a defined format.
In the second step, we use \autoref{prompt:doc2arch2} to generate a list of component names.
Here, we instruct the model to provide only a list of component names without common prefixes and suffixes, like \emph{Component}.
We aim to reduce complexity and facilitate component name retrieval by defining an output format.

\paragraph{Generate Component Names from Source Code}
We also use \CoT prompting to generate the simple \SAM from the source code in two prompting steps.

\begin{prompt}{Code to Architecture (1)}{code2arch1}
    You get the \{Features\} of a software project. Your task is to summarize the \{Features\} w.r.t. the high-level architecture of the system. Try to identify possible components. \{Features\}: \{Content\}
\end{prompt}
\autoref{prompt:code2arch1} queries the model to summarize the features of a software project w.r.t. the system's architecture.
Here, we instruct the model to identify possible components based on the features extracted from the source code.
In this paper, we used "Packages" as the feature.
This feature names all non-empty packages.
We then re-use \autoref{prompt:doc2arch2}.

\paragraph{Generate Component Names from \SAD and Source Code}
Lastly, we consider a combination of the \SAD and the source code to generate the simple \SAM.
Here, we decided to use two modes for the combination.

First, if we extract the \SAM from the documentation and the source code, we can aggregate the results by using \LLMs.
For this purpose, we use the following prompt:
\begin{prompt}{Aggregation}{aggregation}
    You get a list of possible component names. Your task is to aggregate the list and remove duplicates.
    Omit common prefixes and suffixes in the names in camel case.
    \{Output Format (cf. \autoref{prompt:doc2arch2})\}
    Possible component names: \{Possible Component Names\}
\end{prompt}
In \autoref{prompt:aggregation}, we ask the model to aggregate the list of possible component names and remove duplicates.
Afterward, we use the same statements regarding prefixes, suffixes, and output format as in the other prompts.

We calculate the similarity between the component names generated from the documentation and the source code.
To merge the component names, we process them sequentially, starting with the component names from the documentation.
The normalized Levenshtein distance is already known from \TLR tasks \cite{fuchss_establishing_2023,icsa23_ardoco}.
We use this Levenshtein distance \cite{levenshtein1966} to calculate the similarity between the possible component names.
If the similarity is above a certain threshold, we consider the component names as equal and omit the new one.
Since we assume that the \SAD is closer to the actual \SAM than the source code, we start aggregation with the extracted component names from \SAD.
In doing so, we want to complement the component extracted from \SAD with component names extracted from the source code.
Moreover, this also merges similar component names.

\paragraph{Interpretation of Responses}
Among the most challenging aspects of using \LLMs is interpreting the generated responses because there is no guarantee that they adhere to the requested output format.
In \ExArch, the final output of the \LLM should be a list of component names.
This output is determined by the final prompt in each mode (cf. \autoref{prompt:doc2arch2}).
To parse the responses, we consider every line of the final response.
First, we check that the trimmed line starts with the character `-'.
If this is the case, we consider the line as a component name.
We remove the occurrences of 'components' and 'component' from the component names.
Thus, we aim to only get the actual component names.
Third, we remove any spaces from the component's name, as we requested the component names to be in camel case.
After the initial experiments without \ArTEMiS (\cite{fuchss_enabling_2025}), we modified the algorithm to not only remove spaces, but to generate proper camel case by changing the capitalization.
Finally, we remove any duplicates from the list of component names.

\subsection{TransArC Approach}
\label{sec:transarc}
To be able to analyze the effect of LLM-extracted intermediate models on transitive \TLR between \SAD and code, we make use of the \TransArC approach by \citet{keim_transarc_2024}.
\TransArC links \SAD to source code by using \SAMs as intermediate artifacts to bridge the semantic gap.
\TransArC consists of two linking phases: linking documentation to \SAMs, and the models to source code.
The remainder of this section gives a brief overview of them.

In the first phase, \TransArC uses the existing \ArDoCo approach~\cite{icsa23_ardoco} to create trace links between \SAD and \SAMs.
\ArDoCo employs \NLP techniques to analyze the \SAD to identify architectural elements such as components.
\ArDoCo uses various similarity measures and heuristics to link the sentences in the \SAD and components in the \SAM.

In the second phase, \TransArC uses \ArCoTL to establish trace links between \SAMs and the source code.
\ArCoTL first transforms the input artifacts into intermediate representations.
For \SAMs, this includes identifying components, while for code, it involves extracting elements like classes, methods, and packages.
\ArCoTL then applies a series of heuristics to identify correspondences between architectural elements and code entities.
A computational graph combines these heuristics to aggregate the confidence levels of candidates to form the final trace links.

Combining the results from both phases, \TransArC generates transitive trace links between \SAD and source code.
It uses the intermediate \SAMs to enhance the accuracy of trace link recovery, effectively reducing the semantic gap between the documentation and code.
The evaluation of \TransArC (cf. \cite{keim_transarc_2024}), demonstrated its high performance in recovering trace links (weighted average \FOneSc of $0.87$).

We use the \TransArC approach to create trace links between the \SAD and the source code.
Therefore, we use the component names generated by the approach as \SAM to enable trace linking between architecture documentation and source code without needing a manually created component model.

\section{ArTEMiS: An Approach to Extract Architecture Entities from Software Architecture Documentation}
\label{sec:artemis}

In this section, we describe our approach \acs{ArTEMiS} (\acl{ArTEMiS}) that takes \SADs and \SAMs as inputs to identify \TLs between both artifacts. \acused{ArTEMiS}
\ArTEMiS adapts the ideas of two previous approaches of ours: \SWATTR \cite{keim2021_swattr}, which provides trace link recovery between \SAD and \SAM that is also used in \TransArC, and \ExArch, which offers techniques for component identification in text.
Accordingly, \ArTEMiS uses a two-step approach:
It first uses \LLMs to detect architecturally relevant entities in \SADs through a specialized Named Entity Recognition (NER) approach.
Once identified, these entities are matched against actual architecture entities from \SAMs to establish trace links.

\input{figures/artemis_approach}
\autoref{fig:artemis_approach} provides an overview of \ArTEMiS.

For the first step, the identification of architectural entities, \ArTEMiS uses a two-prompt strategy.

The first prompt provides the task description (see \autoref{prompt:artemis_task} for details).
This prompt instructs the \LLM to identify explicitly named software components that are architecturally relevant.
For each component, the \LLM extracts: (1) a primary name, (2) any alternative names or abbreviations, and (3) complete lines where the component appears, either directly or through clear context.
The prompt includes specific rules defining what to include and exclude.
The benefit of this prompt, unlike traditional heuristics or information retrieval approaches, is the capability of \LLMs to identify coreferences and indirect references within their context.
Additionally, \ArTEMiS takes the names of architecture entities from the \SAM as input for this prompt and appends them as positive examples to look out for.
To ensure reliable parsing, the prompt asks the \LLM to output results in structured plain-text format.

The second prompt (see \autoref{prompt:artemis_formatting}) converts the plain-text output into JSON format.
We separated these tasks because combining them into a single prompt frequently produced malformed JSON.
This two-step approach ensures more reliable and parseable results.

After obtaining the JSON output, \ArTEMiS processes the results through several steps.
First, it maps each identified component's mentioned lines back to the original text.
The mapping process begins by searching for exact matches between the mentioned lines and the original text.
If none are found, it calculates string similarity scores to select the best match.
We initially attempted to have the \LLM return line numbers directly, but found that \LLMs struggle with numerical accuracy even when line numbers are provided.
Through this processing, \ArTEMiS produces a structured set of components with their names, alternative names, and line locations.

Finally, \ArTEMiS establishes connections between the \SAM entities and the identified \SAD entities.
The matching process uses both string similarity metrics (Jaro-Winkler \cite{jaroWinkler} and Levenshtein \cite{levenshtein}) and vector embeddings similarity.
For vector embeddings similarity, we use OpenAI's model \enquote{text-embedding-3-large} to embed each of the name and alternative names of the identified named architecture entities from the \SAD on the one hand as well as each name of the \SAM entities on the other hand; we then use cosine similarity to calculate the similarity score between each pair of embedding from the \SAD and embedding from the \SAM.
When one of the similarity scores exceeds a defined threshold, \ArTEMiS creates a link between these entities.
These links enable trace link generation between the various text locations where architectural entities are mentioned and their corresponding model elements.

\ArTEMiS's key advancement is replacing \SWATTR's traditional heuristics with modern \LLMs for architectural entity identification.
For this, \ArTEMiS adopts \ExArch's idea to use \LLMs to identify components in text, but maintains stronger text relationships by tracking both entity occurrences and alternative names throughout the documentation.

\ArTEMiS can be deployed in three different settings.
First, the approach was designed specifically for {\SAD-\SAM \TLR}.
Second, \ArTEMiS can be {integrated into \TransArC} to replace its existing \SAD-\SAM \TLR approach \SWATTR.
Third, \ArTEMiS can be {combined with \ExArch} (using the extracted component names as \SAM) and used {within \TransArC}.

%% file: figures/approach.tex
\begin{figure*}
    \centering
    \begin{tikzpicture}[
            node distance=0.4cm and 0.5cm,
            every node/.style={rectangle, draw, rounded corners, align=center},
            artifact/.style={text width=2.5cm,fill=kit-orange!20},
            process/.style={text width=2cm, fill=white!20},
            process_step/.style={text width=3cm, fill=kit-green!20, draw=kit-green!50, thick},
        ]

        \node[artifact] (doc) {Architecture Documentation};
        \node[artifact] (code) [below=of doc] {Source Code};
        \node[process, right=of code] (features) {Feature Extraction};
        \coordinate (is) at (features.east|-doc);
        \node[process,right=of is,yshift=-0.6cm, inner sep=0pt,minimum size=1.55cm,fill=kit-blue!30] (prompt) {Prompting Strategies};
        \node[artifact] (model) [right=of prompt] {Component Names};
        \node[above=of model, anchor=north, draw=none,xshift=-2cm,yshift=.7cm] (approach_text) {\underline{\acf{ExArch} Approach for Component Name Generation}};

        \begin{pgfonlayer}{background}
            \node[process_step, fit=(doc)(code)(features)(prompt)(model)(approach_text)] (approach) {};
            \draw[->] (doc) -| (model.north);
            \draw[->] (code) -| (model.south);
        \end{pgfonlayer}

        \node[process, fill=kit-purple!20,yshift=-.25cm] (transarc) [below=of code] {TransArC};
        \node[artifact] (tlr) [right=of transarc] {Trace Links};

        \draw[->] (code.south) -- (transarc.north);
        \draw[->] (doc.west) --  ++(-0.5,0)|- (transarc.west);
        \draw[->] ($(model.south) + (0.5cm,0)$) -- ++(0,-0.85cm) -| ($(transarc.north)+(.25cm,0)$);

        \draw[->] (transarc) -- (tlr);
    \end{tikzpicture}
    \caption{Overview of the \acf{ExArch} Approach for \TLR Between \SAD and Code. \ExArch (green box) uses feature extraction and \LLM-based prompting to derive component names from \SAD, source code, or both, producing a simple \SAM (Component Names). \TransArC then uses the \SAD, source code, and the generated component names to recover trace links. Artifacts are shown in orange, prompting in blue, feature extraction in white, and \TransArC in purple.}
    \label{fig:approach}
\end{figure*}

%% file: figures/artemis_approach.tex
\begin{figure*}
    \centering
    \begin{tikzpicture}[
            node distance=1em and 1.5em,
            every node/.style={rectangle, draw, rounded corners, align=center},
            artifact/.style={text width=6em, fill=kit-orange!20},
            process/.style={text width=5em, fill=white!20},
        ]

        \node[artifact] (sam) {\Acl{SAM}};
        \node[artifact, below=2em of sam] (sad) {\Acl{SAD}};

        \node[process, fill=kit-blue!30, text width=6em, right=2em of sad] (ner) {Entity\\Identification\\{\scriptsize(\autoref{prompt:artemis_task})}};
        \node[process, fill=kit-blue!30, text width=5em, right=of ner] (json) {JSON\\Formatting\\{\scriptsize(\autoref{prompt:artemis_formatting})}};

        \node[process, right=of json] (linemap) {Line\\Mapping};
        \node[process, right=of linemap] (match) {Entity\\Matching};

        \node[artifact, right=of match] (tls) {Trace Links\\(SAD--SAM)};

        \draw[->] (sad) -- (ner);
        \draw[->] (ner) -- (json);
        \draw[->] (json) -- (linemap);
        \draw[->] (linemap) -- (match);
        \draw[->] (match) -- (tls);

        \draw[->, dashed] (sam.south) -- ++(0,-0.5em) -| (ner.north);

        \draw[->] (sam.east) -| (match.north);

    \end{tikzpicture}
    \caption{Overview of the \acs{ArTEMiS} Approach for \TLR Between \SAD and \SAM.
    \ArTEMiS takes the \SAD and a \SAM as input.
    It uses two \LLM prompts (blue): the first identifies architecturally relevant entities in the \SAD text, with \SAM entity names provided as examples (dashed arrow); the second converts the plain-text output to structured JSON.
    Line mapping then locates each identified entity in the original \SAD text using exact and similarity-based matching.
    Finally, entity matching links the identified \SAD entities to \SAM entities using string similarity and vector embeddings, producing trace links between \SAD and \SAM.
    Artifacts are shown in orange, heuristic-based steps in white, and \LLM prompting steps in blue.}
    \label{fig:artemis_approach}
\end{figure*}

%% file: sections/eval.tex
\section{Experimental Evaluation}
\label{sec:exp_design}
In this section, we present the evaluation setup (\autoref{para:dataset}), including the dataset, evaluation metrics, and the \LLMs used.
Additionally, we present the baselines to which we compare our approach (\autoref{para:baselines}).
Afterward, \autoref{sec:eval-exarch} discusses the initial experimental evaluation of \ExArch by \citet{fuchss_enabling_2025}.
In \autoref{sec:eval}, we describe our full evaluation of \ArTEMiS and \ExArch.
This section concludes in \autoref{sec:ttv} with our threats to validity.

\subsection{Evaluation Setup: Data and Metrics}
\label{para:dataset}
This section describes the datasets and evaluation metrics that we use for our evaluation.
We use the same setup as in the original \TransArC publication~\cite{keim_transarc_2024}, i.e., we use the same dataset and evaluation metrics to compare the results.
As \ExArch generates all intermediate information needed, we do not rely on manually created \SAMs for configurations that use \ExArch.

\paragraph{Benchmark Dataset}
The \TransArC approach uses a benchmark dataset from \citet{fuchss_establishing_2023} comprising \SAD to \SAM trace links.
\citet{keim_transarc_2024} extended the dataset with trace links between \SAD and source code.
The dataset consists of five open-source projects, each differing in size and domain.
The projects are MediaStore~(MS), TeaStore~(TS), TEAMMATES~(TM), BigBlueButton~(BBB), and JabRef~(JR).
The dataset contains the artifacts themselves and the ground truth for trace links.
\autoref{tab:eval-dataset} provides an overview of the dataset.
Every project has, at most, $14$ components.
The number of source code files ranges from around $100$ to roughly $2000$, and the \SADs comprise 13 up to 198 sentences.
\input{sections/tables/eval-dataset.tex}

\paragraph{Evaluation Metrics}
We use commonly used metrics for \TLR tasks \cite{hayes_advancing_2006,cleland2012}: precision, recall, and their harmonic mean \FOneSc (see \autoref{eq:prec_rec} and \autoref{eq:f1}).
This way, we can compare our approach's performance to the reported results of other state-of-the-art approaches.
\begin{align}
    \text{Precision}\label{eq:prec_rec} & = \frac{\TP}{\TP + \FP}  \text{, } \text{ }  \text{Recall} = \frac{\TP}{\TP + \FN}      \\
    \text{F}_1 \label{eq:f1}            & = 2\times\frac{\text{Precision} \times \text{Recall}}{\text{Precision} + \text{Recall}}
\end{align}

We define the following:
True positives (TPs) are found trace links between the \SAD and the source code that are also contained in the gold standard.
False positives (FPs) are found trace links that are not contained in the gold standard.
False negatives (FNs) are trace links contained in the gold standard but not identified by the approach.

Additionally, we use two different averaging methods.
First, we report the overall (macro) average across all projects without considering their size.
This average provides useful insights into the expected performance on a per-project basis.
Second, we calculate a weighted average based on the number of expected trace links in the gold standard \cite{keim_transarc_2024}.
This weighting offers more in-depth insights into the anticipated effectiveness of the approach for each trace link.

\paragraph{Large Language Models (LLMs)}
We use various \LLMs to generate the simple \SAMs.
We decided to use both closed-source models by OpenAI and locally deployed open-source models.
For OpenAI, we use the following models:
\emph{\GPTFourOMini}, \emph{\GPTFourO}, \emph{\GPTFourTurbo}, \emph{\GPTFour}, and \emph{\GPTThreeFiveTurbo}.
As local models, we use \emph{\CodellamaThirteenB}, \emph{Meta AI \LlamaThreeOneEightB}, and \emph{Meta AI \LlamaThreeOneSeventyB}.

\subsection{Baselines}
\label{para:baselines}
We reuse the baseline approaches of \citeauthor{keim_transarc_2024} for \TransArC \cite{keim_transarc_2024}.
Thus, we can directly compare their results to ours.
The descriptions of the baselines are based on \cite{keim_transarc_2024}.

TAROT \cite{gao_using_2023} and \FTLR \cite{hey_improving_2021} are both recent, state-of-the-art IR-based solutions designed for linking requirements to code.
CodeBERT \cite{feng_codebert_2020} is an \LLM trained to find the most semantically related source code for a given natural language description.
Consequently, all three methods show promising results for similar \TLR problems.

\citet{keim_transarc_2024} also introduced the \ArDoCode approach that uses heuristics to recover trace links between \SAD and source code without using intermediate artifacts, that is based on the \SWATTR approach \cite{keim2021_swattr,icsa23_ardoco}.
They reported that, on average, \ArDoCode was the best-performing approach that does not need \SAMs.
Thus, this approach is one of the important baselines.
Also, we compare to the original \TransArC approach~\cite{keim_transarc_2024} using (manually created) \SAMs.

Moreover, we compare our approaches to the {LiSSA} approach \cite{fuchss_lissa_2025} that uses retrieval-augmented generation and \LLMs to recover \TLs directly.
This approach has been designed to be applicable to various \TLR tasks and was also recently applied to the same dataset.

\subsection{ExArch and TransArC}
\label{sec:eval-exarch}
\input{sections/tables/eval-results-SAD-Code-from-docs.tex}

In this section, we present the results of our evaluation of how to extract component names from \SAD and/or source code using \ExArch and how we integrated it with \TransArC \cite{fuchss_enabling_2025}.

\subsubsection{Extracting Component Names from SAD}
\label{sec:eval-doc}

This section presents the results of our approach \ExArch with \TransArC if we only use the \SAD to generate the component names of a \SAM.
\autoref{tab:eval-results-SAD-Code-from-docs} provides a detailed overview of the results.

The table shows the precision, recall, and \FOneSc for each project and the average values.
The table consists of two sections, one for the baseline approaches and one for our approach using different LLMs.
We highlight the overall best results per project in each section.
In the first section, we took the results of the approaches as presented in the work of \citet{keim_transarc_2024}.
Here, \TransArC uses the manually created \SAM, while the other baseline approaches only use the \SAD and source code.
Overall, the best-performing baselines are \ArDoCode and \TransArC.

The second section of the table presents the results of \ExArch using different \LLMs.
We can see that the performance varies across the projects.
We observe that for the weighted average \FOneSc, the models by OpenAI perform particularly well.
Furthermore, we can see that they perform similarly to \TransArC without the need for manually created \SAMs.
The best model w.r.t. weighted average \FOneSc is \emph{\GPTFourO} with a weighted average \FOneSc of $0.86$ compared to $0.87$ of \TransArC.
The other OpenAI models perform similarly.
According to the classification of \citet{hayes_advancing_2006}, \ExArch \textit{excellently} recovers \TLs between \SAD and code (\GPTFourO).
We highlight that all models outperform the baseline approaches that also do not use manually created \SAMs, and thus, require the same input as \ExArch with \TransArC.

\begin{table}
    \centering
    \caption{Results of a One-Sided Wilcoxon Signed-Rank Test Regarding If \ExArch (\GPTFourO) Using \SAD for Recovery Performs Better Than the Baseline Approaches (Significance Level $\alpha=0.05$,  P-values With * Cannot Be Calculated Exactly.)}
    \label{tab:eval-significance}
    \begin{tabular}{lcccc}
        \toprule
        {Approach / Hypothesis} & {Requires \SAM} & {p-value} & {Significant} \\
        \midrule
        TAROT                          & No                     & .031\phantom{*}  & Yes                  \\
        FTLR                           & No                     & .031\phantom{*}  & Yes                  \\
        CodeBERT                       & No                     & .029*            & Yes                  \\
        ArDoCode                       & No                     & .031\phantom{*}  & Yes                  \\
        LiSSA                          & No                     & .031\phantom{*}  & Yes                  \\
        \TransArC (ours better)        & Yes                    & .970*            & No                   \\
        \TransArC (ours worse)         & Yes                    & .091*            & No                   \\
        \bottomrule
    \end{tabular}
\end{table}
\paragraph*{Significance Tests}
We use Wilcoxon's signed-rank test (one-sided) to calculate the statistical significance of \ExArch's \FOneSc compared to the other baselines.
Since \ExArch works best using \GPTFourO as \LLM, we only compare the results of this configuration to the baselines.
We present the results in \autoref{tab:eval-significance}.
We mark those p-values with an asterisk that cannot be calculated exactly due to ties in the data.
The table shows that \ExArch significantly outperforms all baselines that only use \SAD and code.
Yet, our approach does not outperform \TransArC using manually created \SAM ($p=0.97$).
At the same time, \TransArC is also not significantly outperforming our \LLM-based approach ($p=0.09$).
This shows that \ExArch performs comparably to \TransArC without the need for manually created \SAMs.

\begin{conclusion}{\textbf{Conclusion RQ1.1:}}
    Applying \TransArC with LLM-extracted \SAMs by \ExArch produces similar results as with manually created \SAMs (\ExArch{} (\GPTFourO) weighted avg.\ \FOneSc: $0.86$ vs.\ \TransArC $0.87$; not significantly different at $\alpha=0.05$; see \autoref{tab:eval-results-SAD-Code-from-docs}, \autoref{tab:eval-significance}).
\end{conclusion}
\begin{conclusion}{\textbf{Conclusion RQ1.2:}}
    \ExArch significantly outperforms state-of-the-art \TLR approaches between \SAD and code that do not use \SAMs as intermediates (e.g., \ExArch{} (\GPTFourO) weighted avg.\ \FOneSc $0.86$ vs.\ ArDoCode $0.62$; $p < 0.05$; see \autoref{tab:eval-results-SAD-Code-from-docs}, \autoref{tab:eval-significance}).
\end{conclusion}
\begin{conclusion}{\textbf{Conclusion RQ1.3:}}
    On average, OpenAI's closed-source \LLMs outperform the open-source Llama-based models in this task (best closed-source: \GPTFourO, weighted avg.\ \FOneSc $0.86$; best open-source: \CodellamaThirteenB, $0.71$; see \autoref{tab:eval-results-SAD-Code-from-docs}).
\end{conclusion}

\subsubsection{Generate Component Names from Source Code}
\label{sec:eval-code}
\input{sections/tables/eval-results-SAD-Code-from-code.tex}

This section focuses on \textbf{RQ1.4} and presents the results of \ExArch with \TransArC if we only use the source code to generate the component names.
\autoref{tab:eval-results-SAD-Code-via-code} provides a detailed overview of the results.
As described in \autoref{sec:approach}, we provide a list of all non-empty packages as features for the prompts.
We argue that the packages can be a good representation of the high-level structure of a software project.
Nevertheless, comparing the results to \ExArch using the \SAD to extract the component names, the average performance using only source code is lower.
Especially, the performance for the project \emph{BBB} is bad, i.e., for many models, the \FOneSc is $0$ or close to $0$.
The overall best model is \emph{\LlamaThreeOneSeventyB} with a weighted average \FOneSc of $0.81$.
Notably, this performance is better than its performance when using the \SAD-extracted \SAMs.
Nevertheless, the overall average \FOneSc is worse. 

Since, on average, all other models perform worse than with \SAD, we conclude that the packages alone are insufficient to generate \SAMs for \TLR.
Moreover, this mode's performance is worse than the mode that only considers documentation.
Further, the results vary even more across projects and \LLMs.

\subsubsection{Generate Component Names from SAD and Code}
\label{sec:eval-both}
This section presents the results of \ExArch with \TransArC if we use both the \SAD and the source code to generate the component names of the \SAM.
Here, we consider the two different aggregation strategies described in \autoref{sec:prompts}: the LLM-based strategy and the similarity-based one.
We use a similarity threshold of $t=0.5$ in the experiments for the normalized Levenshtein distance.
Nevertheless, this threshold can also be adjusted to the specific needs of a project.

\input{sections/tables/eval-results-SAD-Code-from-both-combined}

\paragraph{Aggregation via Prompting}
We present our results in \FOneSc regarding the aggregation via prompting in the first part of \autoref{tab:eval-results-SAD-Code-via-both}.
The data shows that the best-performing model is \emph{\GPTFourTurbo} with a weighted average \FOneSc of $0.85$.
The results are slightly worse in macro average, compared to the mode that only uses the \SAD but mostly better than the mode that only uses the source code.

\paragraph{Aggregation via Similarity}
Next, we present the results of our evaluation of the aggregation via similarity.
As described in \autoref{sec:prompts}, we use the normalized Levenshtein distance to aggregate common names.

We present our results in the second part of \autoref{tab:eval-results-SAD-Code-via-both-with-similarity}.
The aggregation is better than the results when only code is used.
The best-performing model is \emph{\GPTFourTurbo} with a weighted average \FOneSc of $0.86$.
Nevertheless, the aggregation via similarity also performs not better than the mode that only uses the \SAD, particularly when considering the non-weighted average \FOneSc.
However, on average, the performance is often better than the aggregation via prompt.

\begin{conclusion}{\textbf{Conclusion RQ1.4:}}
    On average, using only the \SAD to generate the simple \SAM performs best (best avg.\ \FOneSc: \SAD-only $0.76$ (\GPTFourO), code-only $0.58$ (\LlamaThreeOneSeventyB), both-aggregated $0.72$--$0.73$ (\GPTFourTurbo); see \autoref{tab:eval-results-SAD-Code-from-docs}, \autoref{tab:eval-results-SAD-Code-via-code}, \autoref{tab:eval-results-SAD-Code-via-both}).
\end{conclusion}

Overall, we conclude that to extract performing component names for \TLR, the \SAD is a better source than the source code.
\LLMs promise to have great language comprehension capabilities.
Thus, we assume that the task of summarizing and extracting component names based on \SAD is easier for the \LLMs than inferring component names from source code.

\subsubsection{Discussion \& Error Analysis}
\label{sec:discussion}
This section discusses our results and analyzes some errors of the \LLMs.
First, we discuss exemplary differences between the extracted \SAMs from the \SAD using different \LLMs.
Second, we analyze exemplary differences between extracted \SAMs from \SAD and source code.
Finally, we discuss the traceability from the \SAD to the generated \SAM.

\paragraph{Differences between extracted SAMs from SAD using different LLMs}

\input{figures/ms_comparison.tex}

As seen in \autoref{sec:eval-doc}, the performance of the different \LLMs varies across the projects.
To provide insights into the reasons for that, we perform a detailed analysis on the \emph{MediaStore} project with a particular focus on the difference between the results of \emph{\CodellamaThirteenB} and \emph{\GPTFourO}.
We picked this example because they show a rather large gap in precision (\emph{\CodellamaThirteenB} with $0.81$ vs $0.49$ of \emph{\GPTFourO}) while having the same recall.
The first column of \autoref{fig:ms-comparison} shows the \SAM extracted by \emph{\CodellamaThirteenB}, the second column the component names of the original, manually created \SAM, and the third column the \SAM extracted by \emph{\GPTFourO}.
While the \SAMs are not the same, they share similar component names.
The connections between the names indicate that we treat them as being similar.
The highlighting of nodes indicates whether they are matched (green), not matched (red), or related (yellow).
Both generated \SAMs are missing components like \emph{Cache} that are part of the manually created \SAM but not described in the \SAD.
Besides that, the difference is that \emph{\CodellamaThirteenB} generates a component \emph{PersistenceTier} and \emph{\GPTFourO} generates a component called \emph{DataStorage}.
This small difference causes the drop in precision from $0.81$ to $0.49$ in the performance for the \TLR task.
Therefore, we also emphasize that according to model theory, the purpose of a model (e.g., an \SAM) is important \cite{stachowiak1973a}.
\emph{\CodellamaThirteenB} creates component names that are more suitable for the considered \TLR task.
However, this does not mean that they are better for other tasks, and we do not assume that these \SAMs can be directly used in reverse engineering tasks, as these tasks would require more details.

\paragraph{Differences between extracted SAMs from SAD and Code}
In particular, we are interested in the actual differences between the extracted \SAMs from the \SAD and the source code.
Since this requires manual analysis, we only focus on the projects with fewer components according to the manually created \SAMs (cf. \autoref{tab:eval-dataset}), JabRef and TEAMMATES.

\input{figures/jabref_teammates_comparsion}
First, we analyze the differences between the extracted \SAMs for the project \emph{JabRef} using \emph{\LlamaThreeOneSeventyB}.
Our evaluation shows that the performance regarding \TLR is the same as \TransArC using the manually created \SAM.
In \autoref{fig:jabref-comparison}, we show the component names \emph{\LlamaThreeOneSeventyB} extracts.
The first column defines the components extracted using only \SAD, the second column the components of the manually created \SAM, and the third column the components extracted from the source code.
In the third column, we only show the elements that are extracted as main components.
The LLM also generates a list of sub-components with no effects on performance.
We use the same color scheme as in the previous sections.
The names for the matching components of the modes only differ in their letter case, \emph{globals} is missed by both modes, the \SAD-extracted \SAM additionally includes \emph{EventBus}, and the code-extracted adds \emph{Networking}.
The similar results in \TLR suggest that these differences do not matter for the specific \TLR task.

Second, we analyze the differences between the extracted \SAMs for the project \emph{TEAMMATES} using \emph{\GPTFourTurbo}.
Here, our evaluation shows that using \SAD, the performance is comparable to \TransArC, while using the source code, the performance is worse.

In \autoref{fig:teammates-comparison}, we show the component names \emph{\GPTFourTurbo} extracts.
We can see that our extracted \SAM from \SAD is only missing the \emph{GAE Datastore} component.
The extraction from the source code diverges more.
Also, many components are correctly extracted,
some components are abbreviated in the manually created \SAM, while the \LLM extracts the full name from the source code (e.g., \emph{UserInterface}).
However, the \LLM also extracts additional components that are not present in the manually created \SAM.
The component \emph{ArchitectureandMainEntryPoint} originates from the \LLM's output \enquote{Architecture and Main Entry Point}.
Since we instructed the \LLMs to provide component names in camel case, the final component name is as described.
While there are packages \emph{teammates.architecture} and \emph{teammates.main} in the source code, the \LLM could not identify that these packages should not be mapped to a new component.
The first package contains an architecture test and the second package contains the application's main entry point.
The component \emph{TestingandQualityAssurance} is also absent in the manually created \SAM.
While the source code contains packages for test cases, these are not part of the manually created \SAM.
We mark the components that are related to the manually created \SAM with dashed lines.
We assume that the testing-related components extracted via code belong to the \emph{Test Driver} component from the manually created \SAM.
This example shows that if we consider source code for extracting simple \SAMs, distinguishing between code for production and code for testing is challenging.

\paragraph{Traceability from SAD to generated SAM}
To discuss the performance of \ExArch, we briefly analyze the performance of the \TLR approach from \SAD to the generated \SAM.
Since we do not have a \gs for this task, we decided to analyze our results using the following steps:
First, we identify the components generated by the \LLMs.
Second, we manually match the components that are also present in the original \SAM.
Those components are assigned the same unique identifier as in the original \SAM.
All other components are assigned a new unique identifier.
By doing so, we can use the evaluation of \ArDoCo~\cite{icsa23_ardoco} to analyze the \TLR from \SAD to the generated \SAM.
Since we want to do this to get insights into the overall performance from \SAD to source code, we only selected some projects and \LLMs for analysis, based on our evaluation results in \autoref{tab:eval-results-SAD-Code-from-docs}.

We first analyze the \SAD-generated \SAM for \emph{MediaStore} by \emph{\GPTFourO} (see \autoref{fig:ms-comparison}).
We selected this model because the \LLM is the overall best performing. 
Compared to the original \SAM, the generated \SAM contains $11$ components, while the original \SAM contains $14$ components.
It also contains two components that are not present in the original \SAM: \emph{AudioAccess} and \emph{DataStorage}.
Running \ArDoCo with the generated \SAM increases recall for the \TLR between \SAD and \SAM from $0.62$ to $0.79$ compared to running with the manually created \SAM.
The precision decreases from $1.0$ to $0.68$.
Our results for \TLR from \SAD to code also reflect this.

We then analyze the generated \SAM for \emph{BigBlueButton} by \emph{\CodellamaThirteenB}.
We selected this model because it is the best-performing open-source \LLM, except for \emph{BBB}.
Comparing the generated \SAM to the original \SAM, the generated \SAM contains $11$ components, while the original \SAM contains $12$.
The \LLM additionally generates a component called \emph{BigBlueButton}.
Also, it slightly renamed the components \emph{Apps} and \emph{FSESL} to \emph{AppsAkka} and \emph{FSESLAkka}.
The \LLM could not generate the components \emph{BBB web} and \emph{Recording Service}.
It also generated names like \emph{Joiningavoiceconference} that represent tasks but not a component name.
Thus, the recall of the \TLR from \SAD to the generated \SAM is $0.19$ and the precision is $0.27$.
With the original \SAM, the recall is $0.83$ and the precision is $0.88$.
If we consider the package names of the code, one problem can be the generation of the \emph{BigBlueButton} component.
Since all Java packages of the project start with \emph{org.bigbluebutton}, the component can lead to many false-positive trace links that decrease precision.

\subsection{The Impact of ArTEMiS}
\label{sec:eval}
This section covers our evaluation of \ExArch and \ArTEMiS.
First, we describe how we revised the benchmark after the initial experiments of \ExArch \cite{fuchss_enabling_2025}.
Afterward, we discuss the results of our evaluation of \ArTEMiS for \SAD-\SAM \TLR.
Finally, we evaluate \TransArC and \ExArch with the integrated \ArTEMiS approach.

\subsubsection{Revision of the Benchmark}
\input{sections/tables/eval-dataset-v2}
After the initial experiments with \ExArch \cite{fuchss_enabling_2025}, we re-assessed trace links present in the benchmark and created a revised version of it.
We identified some missing trace links (mostly coreferences) and added them.
Also, we identified false positives and removed them.
Finally, we fixed a part of the documentation of \emph{BigBlueButton} where a few sentences were wrongly terminated.
The characteristics of the revised benchmark are summarized in \autoref{tab:eval-dataset-v2}.
The revision of \TLs between \SAD and \SAM also affected the amount of \TLs between \SAD and the code.
We use this revised version of the benchmark for the following evaluations.

\input{sections/tables/eval-results-exarch-v2.tex}

In \autoref{tab:eval-results-SAD-Code-from-docs-v2}, we present the results of our \ExArch approach for the revised benchmark.
We only consider the extraction of component names from the \SAD, as our experiments showed that this is the best-performing mode.
In comparison to the initial version, \ExArch now tries to recover camel-cased component names by removing spaces and enforcing camel-casing in post-processing.
Moreover, we included recent versions of the \LLMs, i.e., \emph{\GPTFourOne} and \emph{\GPTFive}, to provide more insights into the performance of the most recent \LLMs.
As you can see, the results are similar to the initial experiment.
Also, the newer models perform similarly, but the best-performing model is \emph{\GPTFourO} with an average \FOneSc of $0.76$.

\subsubsection{\ArTEMiS for \SAD-\SAM \TLR}
In this section, we present the results of our new \ArTEMiS approach for \TLR between \SAD and \SAM.
To compare our approach with the state of the art, we use the component of \TransArC that creates the \TLs between \SAD and \SAM \textemdash{} namely \SWATTR~\cite{keim_tracelink_2021,icsa23_ardoco}.
We execute \ArTEMiS with three recent \LLMs: \emph{\GPTFourO}, \emph{\GPTFourOne}, and \emph{\GPTFive}.
Since \ArTEMiS does not produce deterministic results, we run each configuration five times and report average results with standard deviation.
\input{sections/tables/eval-artemis.tex}
\autoref{tab:eval-artemis} shows the results of our evaluation.
The best \ArTEMiS configuration w.r.t. \FOneSc uses \emph{\GPTFive}.
On average, this configuration is slightly better than \SWATTR.
Notably, \ArTEMiS achieves a better average recall than \SWATTR.
This is mainly due to the increased recall for MediaStore and TeaStore.
For the other projects, the recall is decreased.
However, the standard deviation is different across the projects and \LLMs.
In particular, for Teammates with \emph{\GPTFive}, the standard deviation of the precision is comparably high with $0.10$.
Still, we can summarize that \ArTEMiS offers an alternative to traditional heuristics-based approaches like \SWATTR for \TLR between \SAD and \SAM.

\begin{conclusion}{\textbf{Conclusion RQ2.1:}}
    \ArTEMiS (\GPTFive) outperforms the state-of-the-art heuristics-based approach \SWATTR in average recall ($0.85$ vs.\ $0.77$) and achieves a slightly higher average \FOneSc ($0.81$ vs.\ $0.80$; see \autoref{tab:eval-artemis}).
\end{conclusion}

\subsubsection{\ArTEMiS and \ExArch in \TransArC}
In this section, we present the results of combining our approaches \ExArch, \ArTEMiS, and \TransArC.
For the \SAD to code \TLR task, we consider three configurations that are compared to the original \TransArC approach and the baseline ArDoCode.
First, the configuration \emph{\TransArC + \ArTEMiS}: Here, we replaced the \SWATTR component of \TransArC that recovers trace links between \SAD and \SAM with our new \ArTEMiS approach using its best \LLM (i.e., \emph{\GPTFive}).
Here, the comparison to the original \TransArC approach is interesting as it also needs existing \SAMs.
Second, the configuration \emph{\TransArC + \ExArch}: This is the same configuration that has been evaluated in the initial experiment \cite{fuchss_enabling_2025}.
Here, we use \ExArch with its best performing \LLM, i.e., \emph{\GPTFourO}, to extract the component names from the \SAD (cf. \autoref{tab:eval-results-SAD-Code-from-docs-v2}).
Third, we analyze the configuration \emph{\TransArC + \ExArch + \ArTEMiS}: This configuration combines both approaches, i.e., we use \ExArch to extract the component names from the \SAD, then use \ArTEMiS to recover the trace links between \SAD and \SAM, and finally use \TransArC to recover the trace links between \SAD and code.
For both configurations, the interesting comparison is mainly ArDoCode, as it is the best baseline that does not require \SAMs for the \SAD to code \TLR task.

\input{sections/tables/eval-complete-combination}
We present the results of our evaluation on \SAD to code \TLR in \autoref{tab:eval-complete}.
Comparing the results of \TransArC with \ArTEMiS to the original \TransArC, we can see that the average recall is increased from $0.77$ to $0.87$.
This is due to the increased recall for MediaStore and TeaStore, analogous to the results of \ArTEMiS vs. \SWATTR.
However, we can also see that the weighted average recall is decreased from $0.93$ to $0.83$, reflecting the decreased recall for projects with more trace links between \SAD and code.
In terms of \FOneSc, we can see that the new \ArTEMiS approach slightly increases the average and weighted average \FOneSc of \TransArC.
Considering the combination of all approaches, i.e., \TransArC with \ExArch and \ArTEMiS, we can see that the results are similar to the results of \TransArC with \ExArch.
This configuration outperforms the baseline approach ArDoCode in terms of \FOneSc.
While the results are similar to the first experiment without \ArTEMiS, we can see some interesting performances.
Notably, the results for JabRef are nearly perfect.
Also, the combination of \ExArch and \ArTEMiS leads to a better \FOneSc in weighted average, meaning that it currently is the best-performing \SAD to code \TLR approach that does not require \SAMs.

\begin{conclusion}{\textbf{Conclusion RQ2.2:}}
    \ArTEMiS (\GPTFive) improves both average recall ($0.87$ vs.\ $0.77$) and average \FOneSc ($0.85$ vs.\ $0.80$) when integrated into \TransArC.
    The combination \TransArC + \ExArch (\GPTFourO) + \ArTEMiS (\GPTFive) achieves the best weighted avg.\ \FOneSc among approaches without \SAMs ($0.87$ vs.\ ArDoCode $0.62$; see \autoref{tab:eval-complete}).
\end{conclusion}

\subsection{\LLMs in Architectural Element Extraction}
\label{sec:discussion}

In our work, we used two \LLM-based approaches, \ExArch and \ArTEMiS, to bridge the semantic gap between \acf{SAD} and code.
We further combine them with \TransArC and enable \SAD-to-code-\TLR without manually created \aclp{SAM}.
In doing so, they collectively answer our overarching research question:

\begin{conclusion}{\textbf{Conclusion RQ0:}}
    Using \LLMs to extract architectural elements yields \TLR performance between \SAD and code comparable to approaches that rely on manually created \SAMs.
    Integrating \ExArch into \TransArC yields a weighted average \FOneSc of 0.86 (vs.\ 0.87 with manual \SAMs), and integrating \ArTEMiS raises average recall from 0.77 to 0.87.
    The combination \TransArC + \ExArch + \ArTEMiS attains a weighted average \FOneSc of 0.87, well above the strongest baseline without \SAMs (ArDoCode, 0.62).
\end{conclusion}

Besides that, the design and evaluation of \ExArch and \ArTEMiS yield broader lessons about using \LLMs in software engineering tasks.
Both \ExArch and \ArTEMiS follow the same design principle: 
They decompose the overall \TLR problem and apply \LLMs only to the subtasks where they excel (\ExArch extracts component names from natural language or code; \ArTEMiS identifies architectural entities in \SAD text) while delegating the final trace link recovery to the heuristics of \TransArC.
Both also adopt a multi-step prompting strategy.
\ExArch uses two prompts per mode: one to elaborate on the architecture, one to produce a rather basic list of names.
A simple list format is easier and more reliable to parse into compared to structured data formats, as minor formatting deviations do not break parsing.
\ArTEMiS requires richer output (entity names, alternative names, and line references), making a structured JSON format necessary but also harder to produce reliably in a single prompt; hence, the second prompt is dedicated solely to converting the plain-text output into JSON.
As \ArTEMiS operates on longer prompts (containing the full \SAD text, the \SAM entity names, and detailed task instructions) and requires a structured JSON formatting step, we run each configuration multiple times and report average results with standard deviation.

Why does \SAD-based extraction work better? We assume that the semantic gap between \SAD and component names is smaller than between code and component names, enabling \LLMs to act like named-entity recognizers on \SAD.
In contrast, component boundaries and names are often implicit in code.
Still, simple \SAMs produced even by smaller \LLMs can be sufficient to enable \TransArC to outperform many state-of-the-art baselines on \SAD-to-code \TLR.
The purpose of these \SAMs matters: the names best for \TLR are not necessarily suited for reverse engineering tasks requiring richer architectural detail.

\subsection{Threats to Validity}
\label{sec:ttv}
In this section, we address potential threats to validity, drawing upon the guidelines provided by \citet{runeson_guidelines_2008}.
We also consider the work of \citet{threats_llms_2024}, which discusses threats specific to experiments involving \LLMs in the context of software engineering.
Unless stated otherwise, the following threats apply to both \ExArch and \ArTEMiS, as well as to their combination with \TransArC.

Regarding \textit{construct validity}, one potential threat is dataset bias.
To mitigate this, we use the same datasets as those employed in state-of-the-art approaches for \TLR.
By doing so, we ensure coverage of various domains and project sizes.
Moreover, our study does not emphasize prompt engineering, meaning we do not extensively modify or compare different prompts.
As a result, the selection of prompts potentially threatens the validity of our findings regarding our research questions.
Another concern is the possible bias in metric selection.
To address this, we employ commonly used metrics in \TLR research.
Overall, we strive to minimize confounding factors that could hinder our ability to address our research questions effectively.

Concerning \textit{internal validity}, there is a risk that other factors influence our evaluation, potentially leading to incorrect interpretations or conclusions.
We adhere to established practices for risk mitigation, utilizing datasets and projects from state-of-the-art approaches.
We also clearly document the origins of these projects and the associated ground truths.
Project quality and consistency variations may still impact the performance of our approaches.
For the combined pipeline (\TransArC + \ExArch + \ArTEMiS), errors may propagate and compound across stages: imprecise component names produced by \ExArch may cause \ArTEMiS to miss or misattribute entity matches, which in turn affects the final \SAD-to-Code trace links.
We mitigate this by evaluating each component individually as well as in combination, allowing us to isolate the contribution of each stage.

There are also potential threats concerning \textit{external validity}.
First, our evaluation is limited to a small number of projects and \TLR tasks, which may affect the generalizability of the results to other projects and tasks.
We utilize an established dataset covering various domains and project sizes to reduce this threat.
However, these datasets mainly consist of open-source projects, which may differ from closed-source projects in key characteristics.
Second, we evaluate the approach using only a limited set of \LLMs, which may limit the robustness of our conclusions.
We further evaluated the approach using only the configuration and the most promising models.
Nonetheless, we evaluated our approaches using various \LLMs to show the performance of fixed prompts using different \LLMs.
A third concern is the use of closed-source models in evaluation.
Since the training data of these models is unknown, we cannot entirely rule out data leakage.
To ensure transparency, we provide our code, prompts, and \LLM responses in our replication packages~\cite{replication,replication_taas25}.
Furthermore, we also use locally deployed models for \ExArch.
Finally, the non-determinism of LLMs poses another threat.
To mitigate this, we set the LLMs' temperature to zero and used the same fixed seed for each run.
By doing so, we make sure that future research can replicate our results.
Setting the temperature to zero is only possible for non-\GPTFive models, as \GPTFive does not allow setting the temperature.
Moreover, for \ArTEMiS, the longer prompts and the structured JSON formatting step can cause non-deterministic behavior with effects on the results even at temperature zero.
Therefore, we run each \ArTEMiS configuration multiple times and report average results with standard deviation to account for this variability.

By conducting our evaluation on established datasets and using widely accepted metrics, we address the \textit{reliability} of our research.
Therefore, we do not introduce a threat regarding the manual creation of the gold standards or datasets.
However, the existing gold standards might be imperfect.

%% file: sections/tables/eval-dataset.tex
\begin{table}
    \centering
    \caption{Number of Artifacts per Artifact Type and Number of Trace Links in the Gold Standard for each Project Based on \cite{keim_transarc_2024}.}
    \label{tab:eval-dataset}
    \begin{tabularx}{\textwidth}{llYYYYY}
        \toprule
        \multicolumn{2}{l}{Artifact Type}                            & MS             & TS & TM  & BBB   & JR            \\
        \midrule
        \Acf{SAD} \cite{fuchss_establishing_2023,keim_transarc_2024} & \# Sentences   & 37 & 43  & 198   & 85    & 13    \\
        \Acf{SAM}                                                    & \# Components  & 14 & 11  & 8     & 12    & 6     \\
        Source Code                                                  & \# Files       & 97 & 205 & 832   & 547   & 1,979 \\
        \midrule
        \SAD-Source Code \cite{keim_transarc_2024}                   & \# Trace links & 50 & 707 & 7,610 & 1,295 & 8,240 \\
        \midrule
        \SAD-\SAM \cite{fuchss_establishing_2023}                    & \# Trace links & 29 & 27  & 51    & 52    & 18    \\
        \bottomrule
    \end{tabularx}
\end{table}

%% file: sections/tables/eval-results-SAD-Code-from-docs.tex
\begin{table*}
    \centering
    \caption{Precision, Recall, and \FOneSc for \TLR Between \SAD and Code (\ExArch: \SAM Derived From \SAD). In the Baselines section, \TransArC uses manually created \SAMs; all other baselines (TAROT, FTLR, CodeBERT, ArDoCode, LiSSA) operate without \SAMs. \ExArch generates component names from \SAD using the listed \LLMs and passes them to \TransArC as simple \SAMs, removing the need for manually created \SAMs. Best result per project within each section is highlighted in bold.}
    \label{tab:eval-results-SAD-Code-from-docs}
    \begin{tabularx}{\textwidth}{cXXXXXXXXXXXXXXX}
        \toprule
                                &                    & \multicolumn{6}{c}{Baselines} & \multicolumn{8}{c}{ExArch}                                                                                                                                                                                                                                                                                                                                                      \\
        \cmidrule(lr){3-8}\cmidrule(lr){9-16}
        Project                 & Met.               & \rotatebox{75}{TAROT}         & \rotatebox{75}{FTLR}       & \rotatebox{75}{CodeBERT} & \rotatebox{75}{ArDoCode} & \rotatebox{75}{LiSSA} & \rotatebox{75}{TransArC} & \rotatebox{75}{GPT-4o mini} & \rotatebox{75}{GPT-4o} & \rotatebox{75}{GPT-4 Turbo} & \rotatebox{75}{GPT-4} & \rotatebox{75}{GPT-3.5 Turbo} & \rotatebox{75}{Codellama 13b} & \rotatebox{75}{Llama 3.1 8b} & \rotatebox{75}{Llama 3.1 70b} \\
        \midrule
        \multirow{3}{*}{MS}     & Pre.               & .09                           & .15                        & .29                      & .05                      & .06                   & \textBF{1.0}             & .49                         & .49                    & .45                         & .49                   & .49                           & \textBF{.81}                  & .49                          & .48                           \\
                                & Rec.               & .24                           & .26                        & .12                      & .66                      & \textBF{.72}          & .52                      & \textBF{.52}                & \textBF{.52}           & .40                         & \textBF{.52}          & \textBF{.52}                  & \textBF{.52}                  & \textBF{.52}                 & .50                           \\
                                & F\textsubscript{1} & .13                           & .19                        & .17                      & .09                      & .11                   & \textBF{.68}             & .50                         & .50                    & .43                         & .50                   & .50                           & \textBF{.63}                  & .50                          & .49                           \\
        \arrayrulecolor{kit-gray30}\midrule\arrayrulecolor{black}
        \multirow{3}{*}{TS}     & Pre.               & .19                           & .19                        & .26                      & .20                      & .34                   & \textBF{1.0}             & .95                         & \textBF{.96}           & \textBF{.96}                & \textBF{.96}          & \textBF{.96}                  & \textBF{.96}                  & .21                          & .62                           \\
                                & Rec.               & .44                           & .25                        & .57                      & \textBF{.74}             & .33                   & .71                      & .67                         & .67                    & .71                         & .71                   & .67                           & .67                           & .71                          & \textBF{.80}                  \\
                                & F\textsubscript{1} & .27                           & .21                        & .36                      & .31                      & .34                   & \textBF{.83}             & .78                         & .79                    & \textBF{.82}                & \textBF{.82}          & .79                           & .79                           & .33                          & .70                           \\
        \arrayrulecolor{kit-gray30}\midrule\arrayrulecolor{black}
        \multirow{3}{*}{TM}     & Pre.               & .06                           & .06                        & .09                      & .37                      & .07                   & \textBF{.71}             & \textBF{.71}                & \textBF{.71}           & \textBF{.71}                & \textBF{.71}          & \textBF{.71}                  & .66                           & .66                          & .62                           \\
                                & Rec.               & .32                           & .30                        & .22                      & \textBF{.92}             & .03                   & .91                      & \textBF{.90}                & \textBF{.90}           & \textBF{.90}                & \textBF{.90}          & \textBF{.90}                  & .49                           & .34                          & .31                           \\
                                & F\textsubscript{1} & .11                           & .10                        & .12                      & .53                      & .04                   & \textBF{.80}             & \textBF{.80}                & \textBF{.80}           & \textBF{.80}                & \textBF{.80}          & \textBF{.80}                  & .56                           & .45                          & .41                           \\
        \arrayrulecolor{kit-gray30}\midrule\arrayrulecolor{black}
        \multirow{3}{*}{BBB}    & Pre.               & .07                           & .04                        & .07                      & .07                      & .20                   & \textBF{.77}             & .71                         & .74                    & .71                         & .63                   & \textBF{.75}                  & .05                           & .73                          & .62                           \\
                                & Rec.               & .18                           & .42                        & .49                      & .57                      & .21                   & \textBF{.91}             & .64                         & .77                    & .66                         & \textBF{.84}          & .54                           & .32                           & .47                          & .43                           \\
                                & F\textsubscript{1} & .10                           & .07                        & .12                      & .13                      & .21                   & \textBF{.84}             & .68                         & \textBF{.75}           & .69                         & .72                   & .63                           & .08                           & .57                          & .51                           \\
        \arrayrulecolor{kit-gray30}\midrule\arrayrulecolor{black}
        \multirow{3}{*}{JR}     & Pre.               & .32                           & .32                        & .49                      & .66                      & .51                   & \textBF{.89}             & \textBF{.89}                & \textBF{.89}           & \textBF{.89}                & \textBF{.89}          & \textBF{.89}                  & \textBF{.89}                  & \textBF{.89}                 & \textBF{.89}                  \\
                                & Rec.               & \textBF{1.0}                  & .93                        & .83                      & \textBF{1.0}             & .01                   & \textBF{1.0}             & \textBF{1.0}                & \textBF{1.0}           & \textBF{1.0}                & .99                   & \textBF{1.0}                  & .99                           & \textBF{1.0}                 & \textBF{1.0}                  \\
                                & F\textsubscript{1} & .49                           & .48                        & .61                      & .80                      & .03                   & \textBF{.94}             & \textBF{.94}                & \textBF{.94}           & \textBF{.94}                & \textBF{.94}          & \textBF{.94}                  & \textBF{.94}                  & \textBF{.94}                 & \textBF{.94}                  \\
        \midrule
        \multirow{3}{*}{Avg}    & Pre.               & .15                           & .15                        & .24                      & .27                      & .23                   & \textBF{.87}             & .75                         & \textBF{.76}           & .75                         & .74                   & \textBF{.76}                  & .67                           & .60                          & .65                           \\
                                & Rec.               & .44                           & .43                        & .45                      & .78                      & .26                   & \textBF{.81}             & .75                         & .77                    & .73                         & \textBF{.79}          & .73                           & .60                           & .61                          & .61                           \\
                                & F\textsubscript{1} & .22                           & .21                        & .28                      & .37                      & .14                   & \textBF{.82}             & .74                         & \textBF{.76}           & .73                         & \textBF{.76}          & .73                           & .60                           & .56                          & .61                           \\
        \arrayrulecolor{kit-gray30}\midrule\arrayrulecolor{black}
        \multirow{3}{*}{w. Avg} & Pre.               & .19                           & .19                        & .28                      & .47                      & .29                   & \textBF{.81}             & .80                         & \textBF{.81}           & .80                         & .80                   & \textBF{.81}                  & .73                           & .75                          & .74                           \\
                                & Rec.               & .63                           & .59                        & .53                      & .92                      & .05                   & \textBF{.94}             & .92                         & \textBF{.93}           & .92                         & \textBF{.93}          & .91                           & .72                           & .67                          & .66                           \\
                                & F\textsubscript{1} & .29                           & .28                        & .36                      & .62                      & .06                   & \textBF{.87}             & .85                         & \textBF{.86}           & \textBF{.86}                & \textBF{.86}          & .85                           & .71                           & .68                          & .68                           \\
        \bottomrule
    \end{tabularx}
\end{table*}

%% file: sections/tables/eval-results-SAD-Code-from-code.tex
\begin{table*}
    \centering
    \caption{\FOneSc for \TLR Between \SAD and Code (\ExArch: \SAM Derived From Source Code). \ExArch generates component names from source code package names using the listed \LLMs and passes them to \TransArC as simple \SAMs. Best result per column is highlighted in bold.}
    \label{tab:eval-results-SAD-Code-via-code}
    \begin{tabularx}{0.75\textwidth}{lZZZZZZZ}
        \toprule
        Approach      & {MS}         & {TS}         & {TM}         & {BBB}        & {JR}         & {Avg.}       & {w. Avg.}    \\
        \midrule
        GPT-4o mini   & .16          & .45          & \textbf{.79} & .23          & .51          & .43          & .60          \\
        GPT-4o        & .28          & .64          & .78          & .12          & .69          & .50          & .68          \\
        GPT-4 Turbo   & \textbf{.55} & .28          & .34          & .00          & \textbf{.94} & .42          & .59          \\
        GPT-4         & .31          & .32          & .61          & .00          & .74          & .40          & .61          \\
        GPT-3.5 Turbo & .22          & \textbf{.68} & .72          & .16          & .32          & .42          & .49          \\
        Codellama 13b & .49          & .00          & .03          & .10          & \textbf{.94} & .31          & .46          \\
        Llama 3.1 8b  & .39          & .45          & .04          & .00          & \textbf{.94} & .37          & .47          \\
        Llama 3.1 70b & .11          & .64          & .75          & \textbf{.46} & \textbf{.94} & \textbf{.58} & \textbf{.81} \\       \bottomrule
    \end{tabularx}
\end{table*}

%% file: sections/tables/eval-results-SAD-Code-from-both-combined.tex
\begin{table}
    \centering
    \caption{\FOneSc for \TLR Between \SAD and Code (\ExArch: \SAM Derived From Both \SAD and Source Code). Component names extracted from both sources are merged using either LLM-based prompt aggregation (top) or Levenshtein-distance-based similarity aggregation ($t=0.5$, bottom). Best result per column and type of aggregation is highlighted in bold.}
    \label{tab:eval-results-SAD-Code-via-both}\label{tab:eval-results-SAD-Code-via-both-with-similarity}
    \begin{tabularx}{.8\textwidth}{lZZZZZZZ}
        \toprule
        Approach      & {MS}         & {TS}         & {TM}         & {BBB}        & {JR}         & {Avg.}       & {w. Avg.}    \\
        \midrule
        \multicolumn{8}{c}{\emph{Prompt-Aggregation}}                                                                          \\
        \midrule
        GPT-4o mini   & .05          & .62          & .79          & .34          & .51          & .46          & .62          \\
        GPT-4o        & .06          & .58          & .78          & .42          & \textbf{.94} & .56          & .82          \\
        GPT-4 Turbo   & .43          & \textbf{.82} & \textbf{.80} & \textbf{.63} & \textbf{.94} & \textbf{.72} & \textbf{.85} \\
        GPT-4         & .31          & .80          & .62          & .53          & \textbf{.94} & .64          & .77          \\
        GPT-3.5 Turbo & \textbf{.44} & .64          & .74          & .44          & .93          & .64          & .80          \\
        Codellama 13b & \textbf{.44} & .56          & .42          & .13          & \textbf{.94} & .50          & .65          \\
        Llama 3.1 8b  & .43          & .00          & .00          & .00          & \textbf{.94} & .27          & .44          \\
        Llama 3.1 70b & .11          & .64          & .71          & .22          & \textbf{.94} & .52          & .78          \\
        \midrule
        \multicolumn{8}{c}{\emph{Similarity-Aggregation} ($t=0.5$)}                                                            \\
        \midrule
        GPT-4o mini   & .27          & .51          & .79          & .49          & .93          & .60          & .82          \\
        GPT-4o        & .28          & .64          & .78          & .54          & .89          & .63          & .81          \\
        GPT-4 Turbo   & .43          & \textbf{.82} & \textbf{.80} & .69          & \textbf{.94} & \textbf{.73} & \textbf{.86} \\
        GPT-4         & .31          & .80          & .61          & \textbf{.70} & \textbf{.94} & .68          & .78          \\
        GPT-3.5 Turbo & .22          & .70          & .72          & .44          & .93          & .60          & .80          \\
        Codellama 13b & \textbf{.63} & .56          & .56          & .10          & \textbf{.94} & .56          & .70          \\
        Llama 3.1 8b  & \textbf{.63} & .34          & .46          & .44          & \textbf{.94} & .56          & .68          \\
        Llama 3.1 70b & .13          & .64          & .78          & .45          & \textbf{.94} & .59          & .82          \\
        \bottomrule
    \end{tabularx}
\end{table}

%% file: figures/ms_comparison.tex
\begin{figure}
    \begin{tikzpicture}[
        every node/.style={rectangle, rounded corners, align=center, minimum height=1.5em,text width = 2.75cm},
        covered/.style={fill=kit-green25},
        uncovered/.style={fill=kit-red25},
        additional/.style={fill=kit-orange25},
        fill fraction/.style={
            path picture={
                \fill[#1]
                (path picture bounding box.north) [sharp corners] rectangle
                (path picture bounding box.south west);
            }
        }]

        \node[text width = 5cm](gs){\textbf{Manual}};

        \node[right = 1em of gs, text width = 3cm](gpt4o){\textbf{\GPTFourO}};

        \node[text width = 3.5cm,draw,uncovered, below=0.5em of gs](audioWatermarking){AudioWatermarking};
        \node[text width = 3.5cm,draw,uncovered, below=0.5em of audioWatermarking](cache){Cache};
        \node[text width = 3.5cm,draw,additional, below=0.5em of cache](db){DB};
        \node[text width = 3.5cm,draw,uncovered, below=0.5em of db](downloadLoadBalancer){DownloadLoadBalancer};
        \node[text width = 3.5cm,draw,covered, below=0.5em of downloadLoadBalancer](facade){Facade};
        \node[text width = 3.5cm,draw,uncovered,fill=kit-orange25,fill fraction=kit-red25, below=0.5em of facade](fileStorage){FileStorage};
        \node[text width = 3.5cm,draw,covered, below=0.5em of fileStorage](mediaAccess){MediaAccess};
        \node[text width = 3.5cm,draw,covered, below=0.5em of mediaAccess](mediaManagement){MediaManagement};
        \node[text width = 3.5cm,draw,covered, below=0.5em of mediaManagement](packaging){Packaging};
        \node[text width = 3.5cm,draw,uncovered, below=0.5em of packaging](parallelWatermarking){ParallelWatermarking};
        \node[text width = 3.5cm,draw,additional, below=0.5em of parallelWatermarking](reencoding){Reencoding};
        \node[text width = 3.5cm,draw,covered, below=0.5em of reencoding](tagWatermarking){TagWatermarking};
        \node[text width = 3.5cm,draw,covered, below=0.5em of tagWatermarking](userDBAdapter){UserDBAdapter};
        \node[text width = 3.5cm,draw,covered, below=0.5em of userDBAdapter](userManagement){UserManagement};

        \node[left = 1em of gs, text width = 3cm](codeLLAMA){\textbf{\CodellamaThirteenB}};
        \node[below=0.5em of codeLLAMA](audioWatermarking1){\vphantom{AudioWatermarking}};
        \node[below=0.5em of audioWatermarking1](cache1){\vphantom{Cache}};
        \node[draw,additional, below=0.5em of cache1](db1){Database};
        \node[below=0.5em of db1](downloadLoadBalancer1){\vphantom{Download} \vphantom{LoadBalancer}};
        \node[draw,covered, below=0.5em of downloadLoadBalancer1](facade1){Facade};
        \node[below=0.5em of facade1](fileStorage1){\vphantom{File Storage}};
        \node[draw,covered, below=0.5em of fileStorage1](mediaAccess1){MediaAccess};
        \node[draw,covered, below=0.5em of mediaAccess1](mediaManagement1){MediaManagement};
        \node[draw,covered, below=0.5em of mediaManagement1](packaging1){Packaging};
        \node[below=0.5em of packaging1](parallelWatermarking1){\vphantom{ParallelWater..}};
        \node[draw,additional, below=0.5em of parallelWatermarking1](reencoder1){ReEncoder};
        \node[draw,covered, below=0.5em of reencoder1](tagWatermarking1){TagWatermarking};
        \node[draw,covered, below=0.5em of tagWatermarking1](userDBAdapter1){UserDBAdapter};
        \node[draw,covered, below=0.5em of userDBAdapter1](userManagement1){UserManagement};
        \node[draw,uncovered, below=0.5em of userManagement1](audioAccess1){AudioAccess};
        \node[draw,uncovered, below=0.5em of audioAccess1](persistenceTier1){PersistenceTier};

        \node[right = 1em of gs, text width = 3cm](gpt4o){\textbf{\GPTFourO}};
        \node[below=0.5em of gpt4o](audioWatermarking2){\vphantom{AudioWater..}};
        \node[below=0.5em of audioWatermarking2](cache2){\vphantom{Cache}};
        \node[draw,additional, below=0.5em of cache2](db2){Database};
        \node[below=0.5em of db2](downloadLoadBalancer2){\vphantom{Download} \vphantom{LoadBalancer}};
        \node[draw,covered, below=0.5em of downloadLoadBalancer2](facade2){Facade};
        \node[draw,additional,below=0.5em of facade2](fileStorage2){DataStorage};
        \node[draw,covered, below=0.5em of fileStorage2](mediaAccess2){MediaAccess};
        \node[draw,covered, below=0.5em of mediaAccess2](mediaManagement2){MediaManagement};
        \node[draw,covered, below=0.5em of mediaManagement2](packaging2){Packaging};
        \node[below=0.5em of packaging2](parallelWatermarking2){\vphantom{ParallelWater..}};
        \node[draw,additional, below=0.5em of parallelWatermarking2](reencoder2){ReEncoder};
        \node[draw,covered, below=0.5em of reencoder2](tagWatermarking2){TagWatermarking};
        \node[draw,covered, below=0.5em of tagWatermarking2](userDBAdapter2){UserDBAdapter};
        \node[draw,covered, below=0.5em of userDBAdapter2](userManagement2){UserManagement};
        \node[draw,uncovered, below=0.5em of userManagement2](audioAccess2){AudioAccess};

        \draw[thick,dashed](db1) -- (db);
        \draw[thick,dashed](db2) -- (db);
        \draw[thick](facade1) -- (facade);
        \draw[thick](facade2) -- (facade);
        \draw[thick,dashed](fileStorage2) -- (fileStorage);
        \draw[thick](mediaAccess1) -- (mediaAccess);
        \draw[thick](mediaAccess2) -- (mediaAccess);
        \draw[thick](mediaManagement1) -- (mediaManagement);
        \draw[thick](mediaManagement2) -- (mediaManagement);
        \draw[thick](packaging1) -- (packaging);
        \draw[thick](packaging2) -- (packaging);
        \draw[thick,dashed](reencoder1) -- (reencoding);
        \draw[thick,dashed](reencoder2) -- (reencoding);
        \draw[thick](tagWatermarking1) -- (tagWatermarking);
        \draw[thick](tagWatermarking2) -- (tagWatermarking);
        \draw[thick](userDBAdapter1) -- (userDBAdapter);
        \draw[thick](userDBAdapter2) -- (userDBAdapter);
        \draw[thick](userManagement1) -- (userManagement);
        \draw[thick](userManagement2) -- (userManagement);

        \begin{pgfonlayer}{background}
            \node[draw,fit=(audioWatermarking)(userManagement)] (manual_bg) {};
            \node[draw,fit=(audioWatermarking1)(persistenceTier1)] (codellama_bg) {};
            \node[draw,fit=(audioWatermarking2)(audioAccess2)] (gpt_bg) {};
        \end{pgfonlayer}
    \end{tikzpicture}
    \caption{Comparison of \SAMs Extracted from \SAD for MediaStore: manually created \SAM (center) vs.\ \ExArch with \CodellamaThirteenB (left) and \GPTFourO (right). Green nodes are exact matches between extracted and manual components (solid lines); orange nodes are approximate matches where names differ but refer to the same concept (dashed lines); red nodes have no match.}
    \label{fig:ms-comparison}
\end{figure}

%% file: figures/jabref_teammates_comparsion.tex
\begin{figure}
\begin{subfigure}[b]{0.49\linewidth}
\resizebox{0.99\linewidth}{!}{
    \begin{tikzpicture}[
        every node/.style={rectangle, rounded corners, align=center, minimum height=1.25em, text width = 2cm},
        covered/.style={fill=kit-green25},
        uncovered/.style={fill=kit-red25},
        additional/.style={fill=kit-orange25},
        fill fraction/.style={
            path picture={
                \fill[#1]
                (path picture bounding box.north)[sharp corners] rectangle
                (path picture bounding box.south west);
            }
        }]
        \node[](gs){\textbf{Manual}};
        \node[draw,fill=kit-red25,fill fraction = kit-green25, below = 0.5em of gs](cli){cli};
        \node[draw,uncovered, below = 0.5em of cli](globals){globals};
        \node[draw,covered, below = 0.5em of globals](gui){gui};
        \node[draw,covered, below = 0.5em of gui](logic){logic};
        \node[draw,covered, below = 0.5em of logic](model){model};
        \node[draw,covered, below = 0.5em of model](preferences){preferences};

        \node[left = 1em of gs, text width = 2.5cm](SAD){\textbf{SAD-extracted}};
        \node[draw,covered, below = 0.5em of SAD](cli2){Cli};
        \node[below = 0.5em of cli2](globals2){\vphantom{globals}};
        \node[draw,covered, below = 0.5em of globals2](gui2){\vphantom{g}Gui};
        \node[draw,covered, below = 0.5em of gui2](logic2){Logic};
        \node[draw,covered, below = 0.5em of logic2](model2){Model};
        \node[draw,covered, below = 0.5em of model2](preferences2){\vphantom{p}Preferences};
        \node[draw,uncovered, below = 0.5em of preferences2](eb){EventBus};

        \draw[thick](cli2) -- (cli);
        \draw[thick](gui2) -- (gui);
        \draw[thick](logic2) -- (logic);
        \draw[thick](model2) -- (model);
        \draw[thick](preferences2) -- (preferences);

        \node[right = 1em of gs, text width = 2.5cm](code){\textbf{Code-extracted}};
        \node[below = 0.5em of code](cli3){\vphantom{cli}};
        \node[below = 0.5em of cli3](globals3){\vphantom{globals}};
        \node[draw,covered, below = 0.5em of globals3](gui3){\vphantom{g}GUI};
        \node[draw,covered, below = 0.5em of gui3](logic3){Logic};
        \node[draw,covered, below = 0.5em of logic3](model3){Model};
        \node[draw,covered, below = 0.5em of model3](preferences3){\vphantom{p}Preferences};
        \node[draw,uncovered, below = 0.5em of preferences3](nw){Networking};

        \draw[thick](gui3) -- (gui);
        \draw[thick](logic3) -- (logic);
        \draw[thick](model3) -- (model);
        \draw[thick](preferences3) -- (preferences);

        \begin{pgfonlayer}{background}
            \node[draw,fit=(cli)(globals)(gui)(logic)(model)(preferences)] (manual_bg) {};
            \node[draw,fit=(cli2)(globals2)(gui2)(logic2)(model2)(preferences2)(eb)] (sad_bg) {};
            \node[draw,fit=(cli3)(globals3)(gui3)(logic3)(model3)(preferences3)(nw)] (code_bg) {};
        \end{pgfonlayer}
        \node[below = 3em of nw] (lowerRight) {}; 
    \end{tikzpicture}
    }
    \caption{JabRef (\LlamaThreeOneSeventyB)}
    \label{fig:jabref-comparison}
\end{subfigure}
\hfill
\begin{subfigure}[b]{0.49\linewidth}
\resizebox{0.99\linewidth}{!}{
    \begin{tikzpicture}[
        every node/.style={rectangle, rounded corners, align=center, minimum height=1.25em, text width = 2cm},
        covered/.style={fill=kit-green25},
        uncovered/.style={fill=kit-red25},
        partly/.style={fill=kit-orange25},
        fill fraction/.style={
            path picture={
                \fill[#1]
                (path picture bounding box.north)[sharp corners] rectangle
                (path picture bounding box.south west);
            }
        }]

        \node[text width = 2.5cm](code){\textbf{Code-extracted}};
        \node[draw,covered,text width = 2.4cm, below = 0.5em of code](client3){Client};
        \node[draw,partly,text width = 2.4cm, below = 2.5em of client3](common3){CommonUtilities};
        \node[draw,partly,text width = 2.4cm, below = 1.4em of common3](lp3){TestingandQual-ityAssurance};
        \node[draw,covered,text width = 2.4cm, below = 0.5em of lp3](logic3){Logic};
        \node[draw,covered,text width = 2.4cm, below = 0.5em of logic3](storage3){Storage};
        \node[draw,partly,text width = 2.4cm, below = 0.5em of storage3](ui3){UserInterface};
        \node[draw,uncovered,text width = 2.4cm, below = 0.5em of ui3](am3){Architectureand\\MainEntryPoint};

        \node[left = 0.9em of code,text width = 2.5cm](gs){\textbf{Manual}};
        \node[draw,covered, below = 0.5em of gs](client){Client};
        \node[draw,uncovered, below = 0.5em of client](gae){GAE Datastore};
        \node[draw, fill=kit-orange25, fill fraction=kit-green25, below = 0.5em of gae](common){Common};
        \node[draw,fill=kit-red25, fill fraction=kit-green25,below = 0.5em of common](ee) {E2E};
        \node[draw,fill=kit-orange25, fill fraction=kit-green25, below = 0.5em of ee](td) {Test Driver};
        \node[draw,covered, left= 1.5em of logic3](logic) {Logic};
        \node[draw,covered, below = 0.5em of logic](storage){Storage};
        \node[draw,fill=kit-orange25, fill fraction=kit-green25, below = 0.5em of storage](ui){UI};

        \node[left = 0.6em of gs,text width = 2.5cm](SAD){\textbf{SAD-extracted}};
        \node[draw,covered, below = 0.5em of SAD](client2){Client};
        \node[below = 0.5em of client2](gae2){\vphantom{GAE Datastore}};
        \node[draw,covered, below = 0.5em of gae2](common2){Common};
        \node[draw,covered](ee2) at (common2|-ee){E2E};
        \node[draw,covered](td2) at (ee2 |- td){TestDriver};
        \node[draw,covered](logic2) at (td2 |- logic){Logic};
        \node[draw,covered, below = 0.5em of logic2](storage2){Storage};
        \node[draw,covered, below = 0.5em of storage2](ui2){UI};

        \draw[thick](client2) -- (client);
        \draw[thick](client3) -- (client);
        \draw[thick](common2) -- (common);
        \draw[thick](ee2) -- (ee);
        \draw[thick](logic2) -- (logic);
        \draw[thick](storage2) -- (storage);
        \draw[thick](td2) -- (td);
        \draw[thick](ui2) -- (ui);

        \draw[dashed,thick](common3) -- (common);
        \draw[dashed,thick](lp3.west) -- (td.east);
        \draw[thick](logic3) -- (logic);
        \draw[thick](storage3) -- (storage);
        \draw[dashed,thick](ui3) -- (ui);

        \begin{pgfonlayer}{background}
            \node[draw,fit=(client)(gae)(common)(ee)(logic)(storage)(td)(ui)] (manual_bg) {};
            \node[draw,fit=(client2)(gae2)(common2)(ee2)(logic2)(storage2)(td2)(ui2)] (sad_bg) {};
            \node[draw,fit=(client3)(common3)(lp3)(logic3)(storage3)(ui3)(am3)] (code_bg) {};
        \end{pgfonlayer}
    \end{tikzpicture}
    }
    \caption{TEAMMATES (\GPTFourTurbo)}
    \label{fig:teammates-comparison}
\end{subfigure}
\caption{Comparison of \SAMs Extracted from \SAD and Source Code vs.\ manually created \SAMs for JabRef (\LlamaThreeOneSeventyB) and TEAMMATES (\GPTFourTurbo). Each subfigure shows the \SAD-extracted \SAM (left), the manually created \SAM (center), and the code-extracted \SAM (right). Green nodes are exact matches (solid lines); orange nodes are approximate matches where names differ but refer to the same concept (dashed lines); red nodes have no match. For JabRef, only the main components are shown in the code-extracted column.}
\end{figure}

%% file: sections/tables/eval-dataset-v2.tex
\begin{table}
    \centering
    \caption{Number of Artifacts per Artifact Type and Number of Trace Links in the revised Gold Standard for each Project. Underline marks the changed values compared to \autoref{tab:eval-dataset}.}
    \label{tab:eval-dataset-v2}
    \begin{tabularx}{\textwidth}{llYYYYY}
        \toprule
        \multicolumn{2}{l}{Artifact Type} & MS             & TS             & TM  & BBB               & JR                                    \\
        \midrule
        \Acf{SAD}                         & \# Sentences   & 37             & 43  & 198               & \underline{87}    & 13                \\
        \Acf{SAM}                         & \# Components  & 14             & 11  & 8                 & 12                & 6                 \\
        Source Code                       & \# Files       & 97             & 205 & {832}             & {547}             & {1,979}           \\
        \midrule
        \SAD-Source Code                  & \# Trace links & \underline{59} & 707 & \underline{8,097} & \underline{1,529} & \underline{8,268} \\
        \midrule
        \SAD-\SAM                         & \# Trace links & \underline{31} & 27  & \underline{57}    & \underline{62}    & 18                \\
        \bottomrule
    \end{tabularx}
\end{table}

%% file: sections/tables/eval-results-exarch-v2.tex
\begin{table*}
    \centering
    \caption{Precision, Recall, and \FOneSc for \TLR Between \SAD and Code with \ExArch on the Revised Benchmark (\SAM Derived From \SAD). Results are shown for three \LLMs (\GPTFourO, \GPTFourOne, \GPTFive). The revised benchmark contains corrected and additional trace links compared to the original (cf.\ \autoref{tab:eval-dataset-v2}). Best result per metric is highlighted in bold.}
    \label{tab:eval-results-SAD-Code-from-docs-v2}
    \begin{tabularx}{.55\textwidth}{llZZZ}
        \toprule
        Project                        & Metric                   & \GPTFourO    & \GPTFourOne  & \GPTFive     \\
        \midrule
        \multirow{3}{*}{MediaStore}    & Precision                & \textBF{.51} & \textBF{.51} & \textBF{.51} \\
                                       & Recall                   & \textBF{.46} & \textBF{.46} & \textBF{.46} \\
                                       & F\textsubscript{1}-score & \textBF{.48} & \textBF{.48} & \textBF{.48} \\
        \arrayrulecolor{kit-gray30}\midrule\arrayrulecolor{black}
        \multirow{3}{*}{TeaStore}      & Precision                & \textBF{.96} & .59          & .62          \\
                                       & Recall                   & .67          & .69          & \textBF{.80} \\
                                       & F\textsubscript{1}-score & \textBF{.79} & .64          & .70          \\
        \arrayrulecolor{kit-gray30}\midrule\arrayrulecolor{black}
        \multirow{3}{*}{Teammates}     & Precision                & \textBF{.75} & .72          & \textBF{.75} \\
                                       & Recall                   & \textBF{.90} & .90          & \textBF{.90} \\
                                       & F\textsubscript{1}-score & \textBF{.82} & .80          & \textBF{.82} \\
        \arrayrulecolor{kit-gray30}\midrule\arrayrulecolor{black}
        \multirow{3}{*}{BigBlueButton} & Precision                & \textBF{.80} & .67          & .70          \\
                                       & Recall                   & .73          & .74          & \textBF{.84} \\
                                       & F\textsubscript{1}-score & \textBF{.77} & .70          & .76          \\
        \arrayrulecolor{kit-gray30}\midrule\arrayrulecolor{black}
        \multirow{3}{*}{JabRef}        & Precision                & \textBF{.89} & \textBF{.89} & \textBF{.89} \\
                                       & Recall                   & \textBF{1.0} & \textBF{1.0} & \textBF{1.0} \\
                                       & F\textsubscript{1}-score & \textBF{.94} & \textBF{.94} & \textBF{.94} \\
        \midrule
        \multirow{3}{*}{Average}       & Precision                & \textBF{.78} & .68          & .69          \\
                                       & Recall                   & .75          & .76          & \textBF{.80} \\
                                       & F\textsubscript{1}-score & \textBF{.76} & .71          & .74          \\
        \arrayrulecolor{kit-gray30}\midrule\arrayrulecolor{black}
        \multirow{3}{*}{w. Average}    & Precision                & \textBF{.83} & .79          & .80          \\
                                       & Recall                   & .92          & .92          & \textBF{.93} \\
                                       & F\textsubscript{1}-score & \textBF{.87} & .85          & .86          \\

        \bottomrule
    \end{tabularx}
\end{table*}

%% file: sections/tables/eval-artemis.tex
\begin{table}
        \centering
        \caption{Comparison of \ArTEMiS to its heuristic-based baseline \SWATTR: Average Precision, Recall, and \FOneSc ($\pm$ Std.\ Dev.) of \SAD to \SAM \TLR over five runs. \ArTEMiS is executed with different \LLMs (\GPTFourO, \GPTFourOne, \GPTFive). Best result per metric is highlighted in bold.}
        \label{tab:eval-artemis}
        \begin{tabularx}{.9\textwidth}{llZZZZ}
                \toprule
                        &           &              & \multicolumn{3}{c}{ArTEMiS}                                                                                     \\
                \cmidrule(lr){4-6}
                Project & Metric    & \SWATTR \small \emph{(baseline)} \normalsize      & \GPTFourO                           & \GPTFourOne                         & \GPTFive                            \\
                \midrule
                \multirow{3}{*}{MediaStore}
                        & Precision & .94          & \textBF{1.0} ($\pm$ .00)            & .99 ($\pm$ .01)                     & .87 ($\pm$ .01)                     \\
                        & Recall    & .55          & .85 ($\pm$ .03)                     & .92 ($\pm$ .02)                     & \textBF{.97} ($\pm$ .02)            \\
                        & \FOneSc   & .69          & .92 ($\pm$ .02)                     & \textBF{.96} ($\pm$ .01)            & .92 ($\pm$ .01)                     \\
                \arrayrulecolor{kit-gray30}\midrule\arrayrulecolor{black}
                \multirow{3}{*}{TeaStore}
                        & Precision & \textBF{1.0} & {.65} ($\pm$ .02)                   & .63 ($\pm$ .02)                     & .59 ($\pm$ .01)                     \\
                        & Recall    & .74          & .86 ($\pm$ .04)                     & .96 ($\pm$ .00)                     & \textBF{.99} ($\pm$ .02)            \\
                        & \FOneSc   & \textBF{.85} & .74 ($\pm$ .01)                     & {.76} ($\pm$ .01)                   & .74 ($\pm$ .01)                     \\
                \arrayrulecolor{kit-gray30}\midrule\arrayrulecolor{black}
                \multirow{3}{*}{Teammates}
                        & Precision & .60          & \textBF{.89} ($\pm$ .05)            & .58 ($\pm$ .07)                     & .78 ($\pm$ .10)                     \\
                        & Recall    & \textBF{.86} & .55 ($\pm$ .04)                     & {.68} ($\pm$ .04)                   & .65 ($\pm$ .07)                     \\
                        & \FOneSc   & \textBF{.71} & .68 ($\pm$ .04)                     & .62 ($\pm$ .04)                     & {.70} ($\pm$ .03)                   \\
                \arrayrulecolor{kit-gray30}\midrule\arrayrulecolor{black}
                \multirow{3}{*}{BigBlueButton}
                        & Precision & \textBF{.90} & .84 ($\pm$ .05)                     & .83 ($\pm$ .04)                     & {.86} ($\pm$ .03)                   \\
                        & Recall    & \textBF{.71} & .37 ($\pm$ .08)                     & {.63} ($\pm$ .06)                   & .62 ($\pm$ .03)                     \\
                        & \FOneSc   & \textBF{.79} & .51 ($\pm$ .09)                     & .71 ($\pm$ .03)                     & {.72} ($\pm$ .02)                   \\
                \arrayrulecolor{kit-gray30}\midrule\arrayrulecolor{black}
                \multirow{3}{*}{JabRef}
                        & Precision & .90          & .93 ($\pm$ .02)                     & .87 ($\pm$ .02)                     & \textBF{.97} ($\pm$ .04)            \\
                        & Recall    & \textBF{1.0} & .90 ($\pm$ .05)                     & \textBF{1.0} ($\pm$ .00)            & \textBF{1.0} ($\pm$ .00)            \\
                        & \FOneSc   & .95          & .91 ($\pm$ .03)                     & .93 ($\pm$ .01)                     & \textBF{.98} ($\pm$ .02)            \\
                \midrule
                \multirow{3}{*}{Average}
                        & Precision & \textBF{.87} & .86 \hphantom{($\pm$ .01)}          & .78 \hphantom{($\pm$ .01)}          & .81 \hphantom{($\pm$ .01)}          \\
                        & Recall    & .77          & .71 \hphantom{($\pm$ .01)}          & .84 \hphantom{($\pm$ .01)}          & \textBF{.85} \hphantom{($\pm$ .01)} \\
                        & \FOneSc   & .80          & .75 \hphantom{($\pm$ .01)}          & .80 \hphantom{($\pm$ .01)}          & \textBF{.81} \hphantom{($\pm$ .01)} \\
                \arrayrulecolor{kit-gray30}\midrule\arrayrulecolor{black}
                \multirow{3}{*}{w. Average}
                        & Precision & .83          & \textBF{.86} \hphantom{($\pm$ .01)} & .76 \hphantom{($\pm$ .01)}          & .81 \hphantom{($\pm$ .01)}          \\
                        & Recall    & .76          & .62 \hphantom{($\pm$ .01)}          & \textBF{.77} \hphantom{($\pm$ .01)} & \textBF{.77} \hphantom{($\pm$ .01)} \\
                        & \FOneSc   & \textBF{.77} & .69 \hphantom{($\pm$ .01)}          & .75 \hphantom{($\pm$ .01)}          & \textBF{.77} \hphantom{($\pm$ .01)} \\
                \bottomrule
        \end{tabularx}
\end{table}

%% file: sections/tables/eval-complete-combination.tex
\begin{table}
    \centering
    \caption{Combining Approaches: Average Precision, Recall, \FOneSc ($\pm$ Std. Dev.) of \SAD to Code \TLR over five runs. TransArC marks the baseline of \TransArC+~\ArTEMiS. ArDoCode is the baseline for \TransArC+~\ExArch and \TransArC+~\ArTEMiS+~\ExArch. We use the best-performing \LLMs for our approaches: For \ExArch, we use \GPTFourO; for \ArTEMiS, we use \GPTFive.}
    \label{tab:eval-complete}
    \begin{tabularx}{\textwidth}{llZZZZZ}
        \toprule
        Project                        & Metric    & \TransArC \small \emph{(baseline)} \normalsize & \TransArC +~\ArTEMiS       & \TransArC +~ \ArTEMiS +~\ExArch & \TransArC \hphantom{+ ArTEMiS} +~\ExArch & ArDoCode \small \emph{(baseline)} \normalsize \\
        \midrule
        Needs \SAMs & & \cmark & \cmark & \xmark & \xmark & \xmark\\
        \midrule
        \multirow{3}{*}{MediaStore}    & Precision & .96       & .86 ($\pm$ .05)            & .53 ($\pm$ .00)                 & .51                                      & .05      \\
                                       & Recall    & .42       & .96 ($\pm$ .03)            & .51 ($\pm$ .00)                 & .46                                      & .59      \\
                                       & \FOneSc   & .59       & .91 ($\pm$ .02)            & .52 ($\pm$ .00)                 & .48                                      & .10      \\
        \arrayrulecolor{kit-gray30}\midrule\arrayrulecolor{black}
        \multirow{3}{*}{TeaStore}      & Precision & 1.0       & .59 ($\pm$ .00)            & .54 ($\pm$ .00)                 & .62                                      & .20      \\
                                       & Recall    & .71       & .99 ($\pm$ .01)            & 1.0 ($\pm$ .00)                 & .80                                      & .74      \\
                                       & \FOneSc   & .83       & .74 ($\pm$ .00)            & .70 ($\pm$ .00)                 & .70                                      & .31      \\
        \arrayrulecolor{kit-gray30}\midrule\arrayrulecolor{black}
        \multirow{3}{*}{Teammates}     & Precision & .75       & .96 ($\pm$ .06)            & .94 ($\pm$ .07)                 & .75                                      & .39      \\
                                       & Recall    & .90       & .67 ($\pm$ .05)            & .68 ($\pm$ .05)                 & .90                                      & .92      \\
                                       & \FOneSc   & .82       & .79 ($\pm$ .03)            & .79 ($\pm$ .02)                 & .82                                      & .55      \\
        \arrayrulecolor{kit-gray30}\midrule\arrayrulecolor{black}
        \multirow{3}{*}{BigBlueButton} & Precision & .82       & .87 ($\pm$ .04)            & .78 ($\pm$ .01)                 & .70                                      & .08      \\
                                       & Recall    & .84       & .74 ($\pm$ .03)            & .64 ($\pm$ .01)                 & .84                                      & .56      \\
                                       & \FOneSc   & .83       & .80 ($\pm$ .02)            & .70 ($\pm$ .00)                 & .76                                      & .14      \\
        \arrayrulecolor{kit-gray30}\midrule\arrayrulecolor{black}
        \multirow{3}{*}{JabRef}        & Precision & .89       & .98 ($\pm$ .04)            & .98 ($\pm$ .04)                 & .89                                      & .66      \\
                                       & Recall    & 1.0       & .99 ($\pm$ .01)            & 1.0 ($\pm$ .00)                 & 1.0                                      & 1.0      \\
                                       & \FOneSc   & .94       & .99 ($\pm$ .02)            & .99 ($\pm$ .02)                 & .94                                      & .80      \\
        \midrule
        \multirow{3}{*}{Average}       & Precision & .88       & .85 \hphantom{($\pm$ .03)} & .75 \hphantom{($\pm$ .03)}      & .69                                      & .28      \\
                                       & Recall    & .77       & .87 \hphantom{($\pm$ .03)} & .77 \hphantom{($\pm$ .03)}      & .80                                      & .76      \\
                                       & \FOneSc   & .80       & .85 \hphantom{($\pm$ .03)} & .74 \hphantom{($\pm$ .03)}      & .74                                      & .38      \\
        \arrayrulecolor{kit-gray30}\midrule\arrayrulecolor{black}
        \multirow{3}{*}{w. Average}    & Precision & .83       & .95 \hphantom{($\pm$ .03)} & .93 \hphantom{($\pm$ .03)}      & .80                                      & .48      \\
                                       & Recall    & .93       & .83 \hphantom{($\pm$ .03)} & .83 \hphantom{($\pm$ .03)}      & .93                                      & .92      \\
                                       & \FOneSc   & .87       & .88 \hphantom{($\pm$ .03)} & .87 \hphantom{($\pm$ .03)}      & .86                                      & .62      \\
        \bottomrule
    \end{tabularx}
\end{table}

%% file: sections/conclusion.tex
\section{Conclusion and Future Work}
\label{sec:conclusion}

We presented two complementary \LLM-based approaches, \ExArch and \ArTEMiS, to bridge the semantic gap between \SAD and code.
\ExArch generates simple \SAMs (component names) from \SAD and/or code to enable \TransArC without manually created \SAMs.
\ArTEMiS recovers \TLs between \SAD and \SAM as a modern alternative to heuristics.
Together with \TransArC, these approaches enable \SAD-to-code-\TLR without manual \SAMs and improve average performance over state-of-the-art baselines.

First, on generating component names, we studied three modes (from \SAD, from code, and both).
Consistent with \autoref{sec:eval-exarch}, generating names from \SAD performs best for the investigated \TLR task.
On the original benchmark, \ExArch with \GPTFourO combined with \TransArC achieves a weighted average \FOneSc of 0.86, comparable to \TransArC with manually created \SAMs (0.87), and significantly outperforms baselines that rely only on \SAD and code.
On the revised benchmark, \ExArch alone (\SAD-only) achieves an average \FOneSc of 0.76 with \GPTFourO.
Extraction from code alone underperforms on average; combining \SAD and code via aggregation helps but does not surpass \SAD-only in macro average.
Across models we evaluated, closed-source OpenAI models on average outperform the open-source Llama-based models.

Second, on linking \SAD to \SAM, \ArTEMiS provides a modern \LLM-based alternative to heuristics.
It improves over \SWATTR in average recall and is slightly better in average \FOneSc (with \GPTFive, recall 0.85 vs. 0.77; \FOneSc 0.81 vs. 0.80).
When integrated into \TransArC, \ArTEMiS increases average recall for \SAD to code \TLR from 0.77 to 0.87 and raises average \FOneSc (0.80 to 0.85).
In contrast, the weighted recall decreases from 0.93 to 0.83 (driven by projects with many links).
The combination \TransArC + \ExArch + \ArTEMiS attains the best weighted average \FOneSc (0.87) among approaches that do not require manually created \SAMs, outperforming the strongest baseline without \SAMs~(ArDoCode,~0.62).

Taken together, these results show that \LLMs are a viable means to extract the architectural elements needed for \TLR between \SAD and code, enabling performance similar to approaches that rely on manually created \SAMs.

In the future, we want to advance code-based \SAM extraction by leveraging architecture recovery techniques; information-fusion methods \cite{RecInfoFusion} appear particularly promising.
We also want to enrich prompts with additional project context (e.g., ADRs, architecture diagrams, and other \SAD sections).
Furthermore, it would be interesting to determine the capabilities of \LLMs regarding trace link identification.
To reduce noise, we aim to separate production and test code when deriving names and refine pre-processing (e.g., package selection and filtering).
We want to explore code summarization to provide more context to \LLMs and study models with integrated \CoT prompting (e.g., \emph{OpenAI o1}).
Finally, we intend to assess per-project calibration (such as similarity thresholds) and devise strategies to improve weighted recall for projects with many links.

%% file: sections/appendix.tex
\section{ArTEMiS Prompt (Task)}
\label{prompt:artemis_task}

\begin{lstlisting}[numbers=left,numberstyle=\footnotesize,basicstyle=\footnotesize,breaklines=true,xleftmargin=2em,backgroundcolor=\color{gray!10},showstringspaces=false,lineskip=-1pt,aboveskip=0pt,belowskip=0pt]
Identify all architecturally relevant software components that are explicitly named in the following text.

For each identified component, provide:
- The primary name (as it appears in the text)
- All alternative names or abbreviations used for the same component in the text (case-insensitive)
- All full lines where the component is mentioned (directly or via clear context)

Rules for identifying components:

1. Only include explicit modular software components with distinct technical responsibilities. These may include:
   - services (e.g., UserService)
   - APIs (e.g., PaymentAPI)
   - adapters, handlers, managers, routers, engines
   - infrastructure components (e.g., Media Server, Presentation Conversion Pipeline)
   - client-side or server-side subsystems (e.g., electron client, backend server)

2. Exclude domain-level entities, even if capitalized - such as business data objects, file types, or general functionalities - unless used as part of a named technical unit.
   Do not include non-technical concepts even if mentioned with verbs like "convert", "generate", or "store" - these are often subject-side actions unless framed as components.

   Examples of domain terms (do not include):
   - image - "Each item includes an image."
   - recommendation - "Recommendations are generated..."
   - file - "Uploads include a JSON file."
   - session - "Each session is stored separately."
   - presentation - "Uploaded presentations go through conversion..."

   Include only when wrapped in named software components that perform active, modular responsibilities (if explicitly named and described).

3. DO include technical subsystems described with proper software roles, and clearly scoped:
   - (Web) server - if described as a component implementing client-server communication or event dispatching
   - (Web) client - if described as rendering or subscribing to events/data
   - Media Server / MS - as a media streaming component implementing SFU/MCU

4. Do not include:
   - Package, class, or namespace names (e.g., common.util, x.y.z)
   - Interfaces (unless directly implemented and deployed)
   - General use of technologies or third-party tools like "React" or "Spring" unless internally wrapped as system components

5. A component is valid if:
   a) Its name includes a functional suffix or architecture-relevant term (Service, Client, Engine, Manager, Adapter, Server, Router, Converter, etc.)
   OR
   b) The text clearly describes it as implementing a technical function within the system (e.g., routing requests, synchronizing state, managing media streams)

6. Reverse pronoun references are allowed only when strongly tied to a previously named component across adjacent lines.
   Do not infer vague or implied components through generic phrases like:
   - it handles the process
   - this system
   - the module

7. Do not create implied components from action nouns (e.g., "conversion", "delivery", "recommendation") unless these are mentioned as named, distinct architectural elements.

8. If an external technology (e.g., MongoDB, Redis, etc.) is used in a custom component (e.g., our RedisPublisher, or MongoSyncService), include that named component - not the technology itself.

Return the results in a clearly structured, unambiguous plain-text format that enables straightforward conversion to JSON (e.g., using key-value sections per component).
\end{lstlisting}

\section{ArTEMiS Prompt (Formatting)}
\label{prompt:artemis_formatting}

\begin{lstlisting}[numbers=left,numberstyle=\footnotesize,basicstyle=\footnotesize,breaklines=true,xleftmargin=2em,backgroundcolor=\color{gray!10},showstringspaces=false,lineskip=-1pt,aboveskip=0pt,belowskip=0pt]
Given the last answer (see below), for each component, return a JSON object containing:
- "name": the primary name of the component (use the most descriptive name).
- "type": "COMPONENT"
- "alternativeNames": a list of alternative or ambiguous names, if applicable.
- "occurrences": a list of lines where the component appears or is referenced.

Output should be a JSON array (and nothing else!), like:
[
    {
        "name": "...",
        "type": "COMPONENT",
        "alternativeNames": [...],
        "occurrences": [...]
    },
    ...
]

Example:
[
    {
        "name": "AuthenticationService",
        "type": "COMPONENT",
        "alternativeNames": ["service"],
        "occurrences": ["The AuthenticationService handles login requests.", "It forwards valid credentials to the UserDatabase.", "The service logs each attempt."]
    },
    {
        "name": "UserDatabase",
        "type": "COMPONENT",
        "alternativeNames": ["DB"],
        "occurrences": ["It forwards valid credentials to the UserDatabase.", "The DB then validates the credentials."]
    }
]
\end{lstlisting}